\documentclass[iop,apj,twocolappendix]{emulateapj}
\usepackage{apjfonts} % Times fonts

\usepackage{graphicx}
\usepackage{verbatim}
\newcommand{\Fermi}{{\it{}Fermi}\ }
\newcommand{\n}{\nodata}
\newcommand{\be}{\begin{itemize}}
\newcommand{\ee}{\end{itemize}}

\def\fermi{\textit{Fermi }}
\def\gr{$\gamma$-ray }
\def\grd{$\gamma$-ray-}
\makeindex
\citeindextrue

\received{July 22, 2011}
\accepted{August 26, 2011}
\journalinfo{Astrophysical~Journal}
\submitted{}
\shorttitle{$\gamma$-Ray and Parsec-scale Jet Properties of Blazars}
\shortauthors{M. L. Lister et al.}

\begin{document}

\title{$\gamma$-Ray and Parsec-Scale Jet Properties of a Complete Sample of Blazars From the MOJAVE Program}
\author{ M. L. Lister\altaffilmark{1},
M. Aller\altaffilmark{2},
H. Aller\altaffilmark{2},
T. Hovatta\altaffilmark{1,9},
K. I. Kellermann\altaffilmark{3},
Y. Y. Kovalev\altaffilmark{4,5},
E. T. Meyer\altaffilmark{6},
A. B. Pushkarev\altaffilmark{7,8,5},
E. Ros\altaffilmark{5,10} 
(for the MOJAVE collaboration)}
\author{
%%%%
M.~Ackermann\altaffilmark{11}
% ...
(for the \textit{Fermi} LAT collaboration)}

\altaffiltext{1}{
Department of Physics, Purdue University, 525 Northwestern Avenue,
West Lafayette, IN 47907, USA;
\email{mlister@purdue.edu}
}
\altaffiltext{2}{
Department of Astronomy, University of Michigan, 817 Dennison Building, Ann Arbor, MI 48 109, USA;
}
\altaffiltext{3}{
National Radio Astronomy Observatory, 520 Edgemont Road, Charlottesville, VA 22903, USA;
%\email{kkellerm@nrao.edu}
}

\altaffiltext{4}{
Astro Space Center of Lebedev Physical Institute,
Profsoyuznaya 84/32, 117997 Moscow, Russia;
%\email{yyk@asc.rssi.ru}
}
\altaffiltext{5}{
Max-Planck-Institut f\"ur Radioastronomie, Auf dem H\"ugel 69,
53121 Bonn, Germany;
%\email{apushkar@mpifr.de}
}
\altaffiltext{6}{Department of Physics and Astronomy, Rice University,
    Houston, TX 77005}

\altaffiltext{7}{
Pulkovo Observatory, Pulkovskoe Chaussee 65/1, 196140 St.
Petersburg, Russia;
}
\altaffiltext{8}{
Crimean Astrophysical Observatory, 98409 Nauchny, Crimea, Ukraine;
}
\altaffiltext{9}{
Owens Valley Radio Observatory, California Institute of Technology}

\altaffiltext{10}{ Departament d'Astronomia i Astrof\'{\i}sica, Universitat de Val\`encia, E-46100 Burjassot, Val\'encia, Spain}
\altaffiltext{11}{W. W. Hansen Experimental Physics Laboratory, Kavli Institute for Particle Astrophysics and Cosmology, Department of Physics and SLAC National Accelerator Laboratory, Stanford University, Stanford, CA 94305, USA}
% ...

\begin{abstract}

We investigate the {\it Fermi} LAT $\gamma$-ray and 15 GHz VLBA radio
properties of a joint $\gamma$-ray- and radio-selected sample of AGNs
obtained during the first 11 months of the {\it Fermi} mission (2008
Aug 4 - 2009 Jul 5). Our sample contains the brightest 173 AGNs in
these bands above declination $-30^\circ$ during this period, and thus
probes the full range of $\gamma$-ray loudness ($\gamma$-ray to radio band
luminosity ratio) in the bright blazar population.  The latter
quantity spans at least four orders of magnitude, reflecting a wide
range of spectral energy distribution (SED) parameters in the bright
blazar population.  The BL Lac objects, however, display a linear
correlation of increasing $\gamma$-ray loudness with synchrotron SED
peak frequency, suggesting a universal SED shape for objects of this
class. The synchrotron self-Compton model is favored for the
$\gamma$-ray emission in these BL Lacs over external seed photon models,
since the latter predict a dependence of Compton dominance on Doppler
factor that would destroy any observed synchrotron SED peak -
$\gamma$-ray loudness correlation. The high-synchrotron peaked (HSP) BL
Lac objects are distinguished by lower than average radio core
brightness temperatures, and none display large radio modulation
indices or high linear core polarization levels. No equivalent trends
are seen for the flat-spectrum radio quasars (FSRQ) in our
sample. Given the association of such properties with relativistic
beaming, we suggest that the HSP BL Lacs have generally lower Doppler
factors than the lower-synchrotron peaked BL Lacs or FSRQs in our
sample.

\end{abstract}
%\tableofcontents
\section{INTRODUCTION} 
\label{s:intro}

The successful launch of the \fermi {\it Gamma-Ray Space Telescope} in
2008 has brought about a new era in our understanding of blazars,
which dominate the extragalactic sky at high energies. Because of
their highly variable fluxes and spectral energy distributions (SEDs),
blazar samples are typically subject to large biases, making it
difficult to study their demographics. With the nearly continuous all-sky
monitoring capabilities of {\it Fermi's} Large Area Telescope (LAT),
however, it is now possible to construct well-defined samples that
can be used to investigate the wide range of jet properties in these
powerful AGNs (e.g., \citealt{1LAC,2009ApJ...707L..56K}).

One of these properties that has been of considerable interest since
the era of the {\it Compton Gamma Ray Observatory} (CGRO) in the
1990s is \gr loudness, or in other words, why only a particular
small subset of known AGNs ($\sim 100$; \citealt{Hartman99}) were
detected by the CGRO's EGRET telescope. Considerable evidence has been
presented by many researchers (e.g.,
\citealt{1995MNRAS.273..583D,2cmPaperIII, 2cmPaperIV, J01, Taylor07})
supporting the idea that relativistic Doppler boosting has a large
impact on AGN \gr emission, but lingering questions regarding the
roles of the flaring duty cycle and the AGN spectral energy distribution
remain. The superior sensitivity and full-time survey operation mode
of \fermi has now provided substantial insight into these issues. With
the release of the 1FGL catalog \citep{1FGL}, the strong impact of SED
characteristics on the fainter \gr AGN population was realized, as the
sky at these levels becomes dominated by high-synchrotron peaked BL
Lac objects. At the same time, the predominant association of \fermi
LAT sources with flat-spectrum radio quasars (FSRQ) and BL Lacs (blazars) has
established Doppler boosting as the primary factor in determining \gr
loudness in the brightest AGNs.

In this paper, we follow up on previous analyses of bright blazars
that were based on the initial 3 month LAT data set presented by
\cite{LBAS}. These studies established several important AGN
radio/$\gamma$-ray connections using quasi-simultaneous VLBA
observations, namely that the \gr photon flux correlates with the
parsec scale radio flux density \citep{MF2, Arshakian2011}, and that
the jets of the LAT-detected blazars have higher-than-average apparent
speeds \citep{MF1}, larger apparent opening angles
\citep{Pushkarev2010}, more compact radio cores \citep{MF2}, strong
polarization near the base of the jet \citep{2011ApJ...726...16L}, and
higher variability Doppler factors \citep{MF4}. In addition, AGN jets
have been found to be in a more active radio state within several months of the
LAT-detection of their strong \gr emission \citep{MF2}, which was
subsequently confirmed by \cite{2010ApJ...722L...7P}.

With the release of the 1st LAT AGN catalog (1LAC; \citealt{1LAC})
based on the initial 11 months of \fermi data, it is now possible to
investigate the impact of Doppler beaming and SED characteristics on
AGN \gr loudness using larger, more complete samples and better
statistics. Here we present a joint analysis of \fermi and VLBA 15 GHz
radio properties of the brightest radio and \gr AGN in the northern
sky, based on data from the LAT instrument, flux density measurements
from the OVRO and UMRAO radio observatories, and the MOJAVE VLBA
program \citep{MOJAVE_V}. In particular we examine the differences in
the SED and \gr properties of BL Lac objects with respect to FSRQs,
and the relative role of relativistic beaming on their \gr loudness.
Several complementary studies will examine the connection between \gr
emission and superluminal speeds (Kadler et al., in prep.), detailed
SED parameters (Chang et al., in prep.), and radio jet activity level
(Lister et al., in prep.).

Throughout this paper, we use a $\Lambda$CDM cosmological model with
$H_0=71$~km~s$^{-1}$~Mpc$^{-1}$, $\Omega_m=0.27$, and
$\Omega_\Lambda=0.73$ \citep{Komatsu09}.

\section{\label{sample} SAMPLE SELECTION}
\subsection{The MOJAVE Survey}

Beginning in 2002, we undertook in anticipation of the {\it Fermi}
mission a program (MOJAVE: \citealt{MOJAVE_I}; \citealt{MOJAVE_VI}) to
assemble the most complete sample possible of bright AGNs that could
be observed relatively easily on a regular basis with the VLBA. This
meant choosing radio sources located in the northern sky that were
bright enough for direct fringe detection on short integration times.
Many of these had been observed regularly for up to seven years by the
preceding VLBA 2 cm Survey program \citep{2cmPaperI}. Because of its
lack of short interferometric baselines, the VLBA effectively filters
out diffuse radio lobe emission, guaranteeing that this sample would
be dominated by AGNs with bright, compact radio cores. As a further
discriminator against steep-spectrum diffuse radio emission, we
carried out the selection at a relatively high radio frequency (15
GHz).

Unlike blazar surveys in the optical or soft X-ray regimes, the radio
emission from the brightest radio-loud blazars is not substantially
obscured by or blended with emission from the host galaxy. Our
VLBA-selected sample thus provides a relatively ``clean'' blazar
sample, namely, one selected solely on the basis of beamed synchrotron
emission from the relativistic jets.

In order to ensure a high overlap with \fermi\ and other blazar
samples, we included in our MOJAVE monitoring program all blazars down
to a specified radio flux density limit. The use of a lower
flux-density cutoff in astronomical surveys is often dictated by
practical concerns such as detector sensitivity or available observing
time, but it also an important parameter in luminosity function
and source population studies. A well-known downside is
the introduction of a luminosity (Malmquist) bias, in which the
average luminosity of sources in the flux-limited sample increases
with redshift. Well-defined flux density limits are essential in blazar
population studies, where the same objects are typically sampled in a
variety of surveys at different wavelengths.  With blazars also comes
the difficulty of substantial flux and spectral
variability. Considerable challenges arise when attempting to compare
data from different wavelength surveys that are not cotemporaneous,
especially when each individual survey may contain or omit certain
objects depending on their activity state at the time the survey was
made.

We addressed the issue of flux variability in MOJAVE by considering a
wide time window during which any source that exceeded the flux limit
was included in the sample. Although this can potentially introduce a
different kind of bias towards highly flaring sources, it has been
effectively used in the 1FGL catalog \citep{1FGL} and in previous
radio blazar surveys (e.g., \citealt{Wehrle1992},
\citealt{Valtaoja1992}). It generally requires a large set of
well-sampled flux density monitoring data. Fortunately we had a large
archive of VLBA (from the 2 cm Survey) and single-dish (from UMRAO and
RATAN) radio flux density measurements of bright AGN ranging from
1994.0 to 2004.0, from which we constructed the original MOJAVE
sample. Any AGN with declination above $-20^\circ$ with measured or
inferred 15 GHz VLBA density that exceeded 1.5 Jy (2 Jy for
declinations $< 0^\circ$) during this period was included (see
\citealt{MOJAVE_V} and the MOJAVE
website\footnote{http://www.physics.purdue.edu/MOJAVE}).  In order to
obtain an even larger overlap with \fermi , we have since extended the
MOJAVE sample to include all sources above 1.5 Jy north of declination
$-30^\circ$ for all epochs from 1994.0 to the present. It is from this
extended survey that we draw the radio-matching sample used in this
paper (\S~\ref{matchingsample}). 

% IFGL compared to 3month: 4x exposure time times 2x fainter sigma = 4 times fainter

\subsection{\label{gamraysample}The 1FM \grd Selected Sample}

In assembling our \gr AGN sample for this paper, our main considerations
were that the sources needed to be suitably bright at \gr 
energies, and have sufficiently strong compact radio emission for
imaging with the VLBA. We also required the sample to be
of reasonable size ($\sim 100$ sources) to ensure good statistics, yet
small enough so that it could still be fully monitored by the MOJAVE
VLBA program. We began by eliminating from the LAT 1FGL catalog
\citep{1FGL} all of the \gr sources known to be associated with
non-extragalactic objects, as well as one gravitationally lensed AGN
(MG J0221+3555 = 1FGL J0221.0+3555). We also excluded five millisecond
\gr pulsars recognized  after the publication of the 1LAC \citep{1LAC}
and 1FGL \citep{1FGL} papers: 1FGL J1231.1$-$1410 \& 1FGL J2214.8+3002 
\citep{Ransom2011}, 1FGL J2017.3+0603 \& 1FGL J2302.8+4443 \citep{Cognard2011},
and 1FGL J2043.2+1709 \citep{2FGL}. 

The specific selection criteria for our initial candidate \grd limited sample were:
\begin{itemize}
\item average integrated $> 0.1$ GeV  energy flux $\ge 3\times 10^{-11}$
erg $\mathrm{cm^{-2}\; s^{-1}}$ between 2008 Aug 4 and 2009 Jul 5. 
\item J2000 declination  $ > -30^\circ$
\item Galactic latitude  $|b| > 10^\circ$
\item not associated with a Galactic source or gravitational lens
\end{itemize}

These criteria yielded a total of 118 candidate AGNs. We note that the
subsequently published 1st LAT AGN Catalog (1LAC; \citealt{1LAC})
listed some additional AGN associations for some 1FGL sources that
were not given in the \cite{1FGL} 1FGL catalog. We used these new
associations to construct our 1FGL-MOJAVE (hereafter 1FM) candidate
list. In the case of three bright \gr sources which had more than one
unique AGN association: 1FGL J0339.2$-$0143, 1FGL J0442.7$-$0019, 1FGL
J1130.2$-$1447, we assumed they were associated with the very bright,
compact FSRQs J0339$-$0146, J0442$-$0017, and J1130$-$1449,
respectively.

For the sky region criteria, we used the position of the radio source
in cases where an AGN association existed, and the LAT position
otherwise. Of the 1FGL sources that met our criteria, only two had no
clear radio source association.  On 2009 Dec 30 and 2009 Dec 31 we
obtained 15 GHz radio telescope pointings at OVRO at the LAT
coordinates of these sources, which yielded 0.11 Jy for 1FGL
J1653.6$-$0158, and an upper limit of 0.01 Jy for 1FGL J2339.7$-$0531.
Since there were numerous possible faint radio counterparts in the LAT
error circle (as seen in NVSS images, \citealt{NVSS}), we dropped
these two LAT sources from the 1FM sample. We subsequently found that all of
the remaining 116 candidate AGNs were bright enough for direct imaging by the
VLBA at 15 GHz (see \S~\ref{vlba_data}). These formed our 1FM
$\gamma$-ray limited sample (Table~\ref{sampletable}).

\begin{figure}
%\epsscale{.85}
%\figurenum{1}
%\plotone{energyflux_vs_radioflux.ps}
\centering
\includegraphics[width=0.5\textwidth,trim=0cm 0cm 0cm 1.7cm,clip]{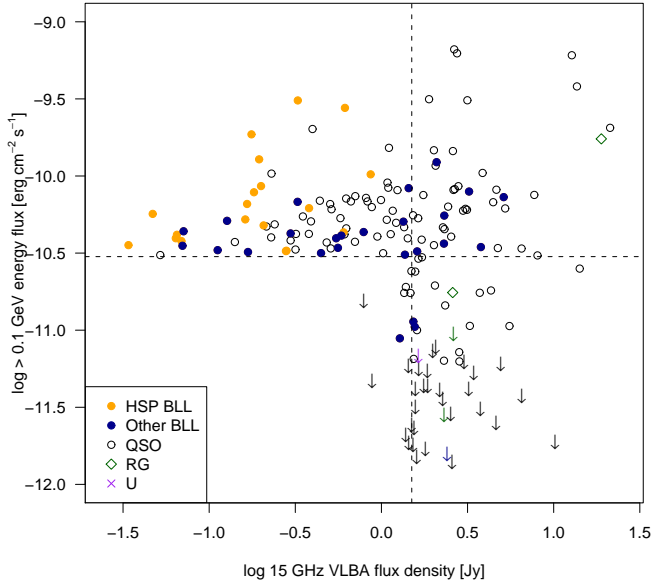}
\caption{\label{energyflux_vs_radioflux} Plot of 11 month {\it Fermi}
  average $>0.1$ GeV energy flux versus 15 GHz VLBA flux density for
  our joint AGN sample. The filled circles represent BL Lac objects,
  with the high synchrotron peaked ones in orange and others in blue.
  The open circles represent quasars, the green diamonds radio
  galaxies, and the purple crosses optically unidentified
  objects. Upper limits on the \gr fluxes are indicated by arrows. All
  of the BL Lac objects are detected by the LAT, with the exception of
  J0006$-$0623. The vertical dashed line indicates the sample radio
  limit of 1.5 Jy, and the horizontal dashed line indicates the \gr
  limit of $3 \times 10^{-11}\; \mathrm{erg \; cm^{-2} \;
    s^{-1}}$. Note that the radio flux density data correspond to
  either a median or ``reference'' epoch coincident with our VLBA
  observations (see \S~\ref{radiofluxes}), and do not necessarily
  coincide with the epoch of maximum radio flux density during the 11
  month LAT period. Some AGNs in the bottom left quadrant thus have
  plotted flux densities below 1.5 Jy.}
\end{figure}

\begin{deluxetable}{lccc}
\tablecolumns{4} 
%\tabletypesize{\scriptsize} 
\tablewidth{0pt}  
\tablecaption{\label{sampletable} AGN Samples}  
\tablehead{\colhead{Sample} & \colhead {$\mathrm{N_{tot}}$} &  
\colhead {$\mathrm{N_{FSRQ}}$} &  \colhead {$\mathrm{N_{BLL}}$}  }
\startdata 
Combined sample        & 173     &  123  &  45 (17)\cr
\grd selected (1FM) &116     &   74  & 41 (17)\cr
Radio-selected     &105     &   86   &  14 (0)\cr
Common to both samples  & 48     &   37    & 10 (0)\cr
\enddata
\tablecomments{$\mathrm{N_{tot}}$ = total number of AGNs,\\
$\mathrm{N_{FSRQ}}$ = total number of flat-spectrum radio quasars,\\
$\mathrm{N_{BLL}}$ = total number of BL Lac objects (number of which are known to be high-spectral peaked). \\
}
\end{deluxetable}

\subsection{\label{matchingsample}The 1FM-Matching Radio-Selected Sample}
For the purposes of constructing a matching radio-selected sample, we
used the same sky region criteria as the 1FM, this time choosing all AGNs
known to have exceeded $S_\mathrm{VLBA}$ = 1.5 Jy at 15 GHz during the initial
\fermi 11 month period, without regards to \gr flux. To carry out this
selection, we relied on MOJAVE VLBA measurements, as well as OVRO and
UMRAO single-dish data, from which compact (VLBA) flux densities could
be estimated (\S\ref{radiofluxes}).  There are 105 AGNs in our final 1FM matching
radio-selected sample, 48 of which are also in the 1FM \grd selected
sample.  In Figure~\ref{energyflux_vs_radioflux} we plot the 11-month
$>0.1$ GeV average \gr energy flux versus 15 GHz VLBA flux
density, which shows the region of the flux-flux density plane
covered by our survey. We note that the radio flux density data plotted in
Figure~\ref{energyflux_vs_radioflux} correspond to either a median or
``reference'' epoch coincident with our VLBA observations (see
\S~\ref{radiofluxes}), and do not necessarily coincide with the epoch
of maximum radio flux density during the 11 month LAT period. Thus, some
AGNs in the radio-selected sample have plotted flux densities
below 1.5 Jy.

\subsection{Selection Biases}

We have assembled two complete samples of the brightest AGNs in the
northern \gr and radio sky, as seen during the first 11 months of the
\fermi mission. We list their general properties in
Table~\ref{1gentable}. The optical redshifts and classifications are
from the compilations of \cite{MOJAVE_V} and NED (see Appendix for
notes on individual sources). Note that we classify J0238+1636 as a
quasar because of its occasional broad emission lines \citep{Raiteri2007},
and the presence of a break in its \gr spectrum that is characteristic
of FSRQs \citep{abdo_sedbreak}. For the purposes of this paper, we
have  grouped two narrow-line Seyfert 1 galaxies J0948+0022 and
J1504+1029 \citep{2011nlsg.confE..24F} with the quasar class.

The SED data are taken mainly from
\cite{Chang_thesis}, \cite{abdo_sed}, and other papers in the
literature as indicated in column 8. We use the following nomenclature
for high-, intermediate- and low-synchrotron peaked blazars: LSP $<
10^{14}$, $10^{14} <$ ISP $< 10^{15}$, and HSP $> 10^{15}$, where the
values refer to the synchrotron SED peak frequency ${\nu_s}$ in Hz.

Although our \gr and radio selections are both made on the basis of
compact beamed jet emission, there is only a 28\% overlap in the two
samples. This is perhaps lower than might be expected, given the
strong correlations previously seen between the 1LAC catalog and
flat-spectrum radio sources \citep{1LAC}. As we discuss in
\S~\ref{gamrayloudness}, however, this is mainly a consequence of the
wide range of \gr loudness in the bright blazar population. There is
also some likelihood that any particular AGN will not have a LAT
association because it happens to lie in a confused region that
contains several bright \gr sources, or has a high diffuse \gr
background. The latter case is less likely to occur however for the
bright non-Galactic-plane sources we are considering. We have
carefully examined our candidate list and found only one possible case
of a missed association: 1FGL J1642.5+3947.  Recent analysis by the
LAT team \citep{2010arXiv1012.2820S} has led us to associate this
source with the FSRQ J1642+3948 (3C~345). 

The nature of our \gr sample selection differs from that of our radio
sample, since it uses average fluxes instead of maximum measured flux
densities, and it spans a wide energy band compared to the radio. It
is thus more sensitive to the shapes of the AGN SEDs, which can have
curvature and breaks within the LAT detector band. The spectral
response function of the LAT detector and its favoritism towards
harder sources causes some selection bias towards faint
high-synchrotron peaked AGNs \citep{1FGL}.  We note, however, that the
sources in our 1FM sample are selected well above the instrument
sensitivity level of the LAT detector, and should be devoid of biases
related to threshold effects.

The above selection biases do not have a large impact on the analysis
presented in this paper, since our primary goal is to identify broad
statistical trends between the \gr emission and radio jet
properties. For this purpose a representative blazar sample that spans
a wide range of SED peak frequency and \gr loudness is appropriate.
Future studies using more extensive \fermi data will address these
issues in considerably more detail, with better statistics. These will
be needed for accurate determination of the blazar \gr luminosity
function for different redshift ranges and optical sub-classes.

\section{OBSERVATIONAL DATA}

\subsection{\label{radiofluxes}Radio Flux Density Data}

We list the radio flux density data for our sample in
Table~\ref{2fluxtable}.  For each AGN we selected a VLBA ``reference''
epoch, which was chosen to be the closest MOJAVE VLBA observation to
the end of the initial 11-month \fermi period. In the case of 41
sources, no VLBA data were available within this period, so we used
the first available MOJAVE VLBA epoch following this period.  The latter
epoch dates ranged from 2009 July 23 to 2010 Nov 29. We list the reference epoch
dates and total 15 GHz VLBA flux densities in columns 3 and 4,
respectively.  In column 5 we list the median single dish flux density
from OVRO at 15 GHz (or 14.5 GHz at UMRAO as indicated) during the
same 11 month period \citep{Richards11,UMRAO}.

The vast majority of the radio sources in our sample are strongly core
dominated at 15 GHz \citep{MOJAVE_V}, and therefore there is typically
very little flux density that is missed by the VLBA. In order to
estimate this amount for each source, we compared our historical
MOJAVE flux density measurements with cotemporeaneous 14.5 GHz UMRAO
measurements (within 7 days), and 15 GHz OVRO measurements that were
interpolated to the VLBA epoch date. By taking the mean of these
single dish-minus-VLBA flux density measurements, we obtained the
extended flux density values that are tabulated in column 6.  For the
sources with no value listed, the amount of extended flux density was
smaller than 3 times the associated measurement error. The errors in
our VLBA flux density measurements are on the order of 5\%, while the
single-dish errors are smaller \citep{Richards11, UMRAO}.

For the purposes of determining an average \gr loudness parameter $G_r$
for each source during the first 11 months of LAT science operations
(Section~\ref{gamrayloudness}), we required an estimate of the median
15 GHz VLBA radio flux density during the initial 11-month \fermi
period.  Since the single dish radio monitoring data were much more
densely sampled than the VLBA data, we estimated the latter by using
the single dish median in column 5 of Table~\ref{2fluxtable} and
subtracting the source's extended flux density (assuming zero extended
flux density for those sources with no value listed in column 6). For
28 sources which lacked a single dish median value, we used the VLBA flux
density at the reference epoch (column 4).

We also collected radio variability statistics for 84\% of our
AGN sample using 15 GHz OVRO observatory data taken during the first
11 months of the \fermi mission. The modulation index data are
described and tabulated by \cite{Richards11}. This index is defined
as the standard deviation of the flux density measurements in units of
the mean measured flux density (e.g., \citealt{Quirrenbach2000}), and
is less sensitive to outlier data points than other variability measures. 

\subsection{\label{vlba_data}VLBA Data}

The 15 GHz radio VLBA data were obtained as part of the MOJAVE
observing program \citep{MOJAVE_V}, and consist of linear polarization
and total intensity images with a typical image FWHM restoring beam of
approximately 1 milliarcsecond. This corresponds to a scale of a few
parsecs at the typical redshifts ($z \simeq 1$) of our sample AGNs. We obtained
fractional linear polarization and electric vector position angle
measurements for the reference epoch image using the methods described
by \cite{MOJAVE_I}.  We calculated the mean position angle of each jet
on the sky by taking a flux density-weighted average of the position
angles of all Gaussian jet components fit to all available 15 GHz VLBA
epochs up to the end of 2010 in the MOJAVE archive. A description of
the Gaussian model fitting method is given by \cite{MOJAVE_VI}. We
used the Gaussian fit to the flat-spectrum core component of each jet
at the VLBA reference epoch to determine a rest frame core brightness
temperature $T_b$ (column 5 of Table~\ref{3jettable}) for each jet
according to
\begin{equation}
T_b= 1.222\times10^{12}\; { S_\mathrm{core}\; (1+z) \over \nu^{2}\; \theta_\mathrm{maj}\; \theta_\mathrm{min}} \quad \qquad \mathrm{K},
\end{equation}
where $S_\mathrm{core}$ is the fitted core flux density in Janskys at $\nu =
15$ GHz, and $\theta_\mathrm{maj}$ and $\theta_\mathrm{min}$ are the
FWHM dimensions of the fitted elliptical Gaussian core components
along the major and minor axes, respectively, in milliarcseconds. In cases where the
best fit to the core was a zero-size (point) component, we used the
signal-to-noise ratio formula of \cite{2cmPaperIV} to determine a
lower limit on $T_b$. For the 26 sources without a redshift we
assumed $z=0.3$ in calculating $T_b$ (and $G_r$ in
Section~\ref{gamrayloudness}), since most of these are BL Lac objects,
and this corresponds to the median BL Lac redshift in our sample.

We obtained pc-scale jet opening angle measurements (as projected on
the sky) using the method described by \cite{Pushkarev2010}. We used a
stacked image of all available 15 GHz epochs in the MOJAVE archive for
this purpose.  The median opening angle value for each jet is listed
in Table~\ref{3jettable}. Five \grd selected sources with weak radio
flux densities ($<$ 200 mJy) did not possess sufficiently bright jet
emission to estimate their opening angles. These were J0136+3906,
J0507+6737, J1037+5711, J1303+2433, and J1725+1152.  Additionally, the
FSRQ J0957+5522 (4C +55.17) is largely resolved by the the long baselines of the
VLBA at 15 GHz and thus has a low brightness temperature and very
little measurable jet structure
\citep{McConville,2005A&A...434..449R}. Our opening angle measurements
based on the stacked-epoch images are in generally good agreement with
the single-epoch measurements of the same sources by
\cite{Pushkarev2010}.  In some sources our measured opening angle was
much wider, because of the presence of low-brightness jet emission that
was below the noise level in the single-epoch image.  In a few other
cases, the ejections of new moving jet features along different
position angles over time resulted in a wider apparent opening angle
than seen in the single-epoch image.

\subsection{\gr Loudness \label{gamrayloudness}}
Our chosen statistic for describing \gr loudness is the ratio of
average \gr luminosity during the first 11 months of the \fermi
mission to the median 15 GHz VLBA radio luminosity.  We have compiled
this ratio $G_r$ for all the AGNs in our sample using the 1FGL $>$ 0.1
GeV \gr energy flux measurements of \cite{1FGL} and the radio data
described in \S~\ref{radiofluxes}.  These ratios are listed in
Table~\ref{2fluxtable}.  

In the 1FGL catalog, the $\gamma$-ray source significance is measured
in terms of the Test Statistic (TS), where TS is defined as 2 times
the difference in the log(likelihood) measure with and without the
source included \citep{1996ApJ...461..396M}.  All sources in the 1FGL
and 1LAC catalogs have TS $> 25$. For the 1FM radio-matching sources
that had no associations in the 1LAC catalog, we determined an upper
limit on the $>0.1$ GeV photon flux directly from the 11-month \fermi
LAT data, assuming a point source with a power law spectrum. We
analyzed photons of the `diffuse'' class with a zenith angle smaller
than 105$^{\circ}$ in the energy range 0.1--100\,GeV within a circular
region of interest (RoI) with a radius of 12$^{\circ}$ centered around
the radio position of the source. We modeled the $\gamma$-ray emission
from the RoI using extended Galactic and isotropic templates and all
sources from the 1FGL catalog.  We let the model parameters of sources
in the RoI vary, and froze those of the outer sources to the catalog
values.  We used the standard \textsl{Fermi}-LAT {\it ScienceTools}
software package (version v9r16p1) with the instrument response
functions `P6\_V3\_DIFFUSE'' to obtain a flux value for each source.
To obtain the upper limits we increased the flux from the
maximum-likelihood value until $2\Delta \log(\mathrm{likelihood}) = 4$
\citep{2005NIMPA.551..493R}. Our final upper limits thus correspond to
$\sim2 \sigma$.  For sources with $\mathrm{TS}<1$ we calculated a 95\%
upper limit using a Bayesian approach \citep{Helene1983319}. We
converted these to energy fluxes according to
\begin{equation}
\label{gamlumeqn} 
S_{0.1} = {(\Gamma -1) \; C_1\;  E_1\;F_{0.1} \over (\Gamma -
  2)} \left[1- {\left(E_1 \over
    E_2\right)}^{\Gamma-2}\right] \qquad \mathrm{erg \; cm^{-2} \; s^{-1}},
\end{equation}
where $F_{0.1}$ is the upper limit on the photon flux above $E_1$ =
0.1 GeV in $\mathrm{photons \; cm^{-2} \; s^{-1}}$, $E_2$ = 100 GeV,
and $C_1 = 1.602 \times 10^{-3} $ erg $\mathrm{GeV^{-1}}$. In calculating these
upper limits, we fixed the photon spectral index to $\Gamma = 2.1$.

We converted the measured energy fluxes and upper limits to \gr
luminosities according to 
\begin{equation}
L_\gamma = {4\pi D_L^2 \;  S_{0.1} \over (1+z)^{2-\Gamma} }  \qquad \mathrm{erg \; s^{-1}},
\end{equation}
where $D_L$ is the luminosity distance in cm, $ \Gamma$ is the
11-month average \gr photon spectral index for sources with 1LAC associations and $\Gamma = 2.1$ otherwise, $z$ is
the redshift, and $S_{0.1}$ is the 11-month average energy flux (or upper limit) above 0.1 GeV in $\mathrm{erg \; cm^{-2} \; s^{-1}}$.

 As discussed by \cite{1FGL}, the lower-energy LAT band photon fluxes
 are poorly determined; therefore the energy flux over the full
 band is better defined than the 0.1 to 100 GeV photon flux. The
 average 11-month energy fluxes tabulated by \cite{1FGL} were found by summing
 the energy fluxes in five individual bands over this energy range.

We calculated the radio luminosities over a 15 GHz wide bandwidth
according to:
\begin{equation} 
L_R = {4\pi \; D_L^2 \; \nu \; S_{\nu} \over (1+z)} \qquad \mathrm{erg \; s^{-1} ,}
\end{equation}
where $S_\nu$ is the median VLBA flux density at $\nu = 15$ GHz as
defined in \S~\ref{radiofluxes}. We assumed a flat radio spectral index ($\alpha = 0$) for the purposes of
the $k$-correction and luminosity calculations.

% Fermi LAT 11 month covers 2008-08-04 to 2009-07-04 (Abdo 2010 ApJs 188, 405)

\section{DATA ANALYSIS AND DISCUSSION}

\subsection{Redshift Distributions}

\begin{figure}[t]
%\epsscale{.95}
%\figurenum{1}
%\plotone
\centering
\includegraphics[width=0.47\textwidth,trim=0cm 0cm 0cm 0cm,clip]{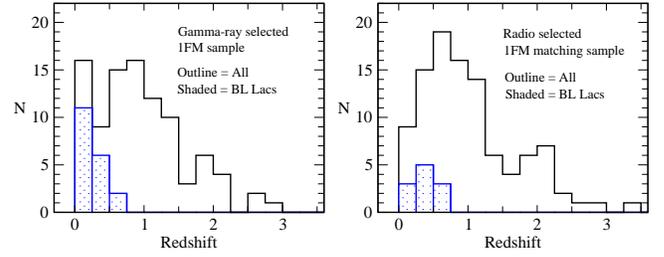}
\caption{\label{zhistograms} Left panel: redshift
  distribution of the \grd selected 1FM sample. The full sample is
  represented by the solid line, and the BL Lac objects are
  shaded. There is one radio galaxy (J0319+4130 = 3C 84) at $z = 0.0176$. Right panel:
  redshift distribution for the radio-selected 1FM matching sample.
  There are four radio galaxies in the sample, all in the first ($z <
  0.25$) bin.}
\end{figure}

The redshift data on our AGNs are incomplete (see Appendix), with
missing values for 4 sources in the radio-selected sample, and 22
sources in the \grd selected sample (the sources J0050$-$0929 and
J0818+4222 are common to both samples).  In Figure~\ref{zhistograms}
we plot the redshift distributions for our samples. The redshifts
range from $z = 0.00436$ to $z = 3.396$, and the distributions are
generally peaked between $z=0.5$ and $z=1$. Kolmogorov-Smirnov (K-S)
tests do not reject the null hypothesis that the \grd selected and
radio-selected samples are drawn from the same parent redshift
distribution, even when the sources in common to both samples are
excluded (D = 0.20, probability = 0.27). We find no statistical
differences in the redshift distributions of the non-LAT detected
versus LAT-detected AGNs in the combined samples (D = 0.16,
probability = 0.49).

With respect to the redshifts of the quasars in the two samples, the
K-S test suggests a marginal statistical difference in their
distributions (D = 0.079, probability = 0.96). There are an
insufficient number of radio galaxies to perform any statistical tests
on them (there are 4 in the radio-selected sample; one of these is
also in the \grd selected sample). The overall redshift distribution
of the \grd selected sample has an additional peak at low redshift,
due to the presence of at least 9 high synchrotron peaked (HSP) BL
Lacs that are not in the radio selected sample (8 additional HSP BL
Lacs lack redshift information). These objects also bring the overall
fraction of BL Lacs up to 35\% in the \grd selected sample, as
compared to only 13\% for the radio-selected sample.

%Given the similar wide redshift range of both samples, we conclude that there is no large-scale attenuation of the \fermi \gr flux from IGM scattering/pair-production processes, in comparison
%to the radio flux.

Because of the similarities in the properties and redshift
distributions of the \gr and radio-selected samples, for the remainder
of this paper we will no longer distinguish between them, referring
instead to the joint sample of 173 AGNs. 

\subsection{\gr Loudness and Synchrotron Peak Frequency}

\begin{figure}[t]
%\epsscale{.85}
%\figurenum{1}
%\plotone{lgam_vs_lradio.ps}
\centering
\includegraphics[width=0.5\textwidth,trim=0cm 0cm 0cm 2cm,clip]{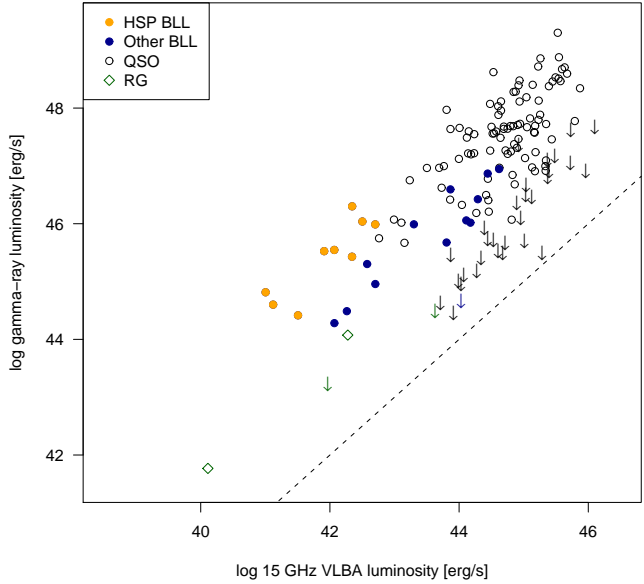}
\caption{\label{lgam_vs_lradio} Plot of average \gr luminosity versus
  median VLBA 15 GHz radio luminosity.  The filled circles represent
  BL Lac objects, with the high synchrotron peaked ones in orange and
  others in blue. The open circles represent quasars and the green
  diamonds radio galaxies. The arrows represent upper limits based on
  the 11-month LAT data.  The dashed
  line represents the 1:1 luminosity ratio line. }
\end{figure}

A primary goal of our study is to examine the range of \gr loudness
($G_r$) present in the bright blazar population, and its dependence on
other AGN jet properties.  Since we have obtained data in several
complete regions of the \grd radio plane (i.e., \grd
bright/radio-faint; \grd faint/radio-bright; \grd bright/radio-bright)
we can be assured of sampling the largest possible range of $G_r$ in
the brightest northern-sky blazars. Future studies of the \grd
weak/radio-weak region will be important for verifying whether the
trends we identify here extend to the fainter blazar population.

In Figure~\ref{lgam_vs_lradio} we plot \gr luminosity against 15 GHz
VLBA luminosity.  Despite our use of an average \gr luminosity over an
11-month period, the linear relationship for the non-censored data has
only moderate scatter (0.6 dex). A linear regression fit to the
non-censored data yields $\log{L_{\gamma}} = (0.92 \pm 0.05) \log{L_R}
+ 6.4 \pm 2$.  The $G_r$ values, which reflect the perpendicular distance of the
data points from the dashed 1:1 line, span nearly 4 orders of
magnitude, from below 3 to $\sim 15000$. A clear division between the
HSP and lower-synchrotron-peaked BL Lacs is evident, with the former
having higher \gr loudness ratios.

\begin{figure}[t]
%\epsscale{.75}
%\figurenum{1}
%\plotone{rg_histograms.ps}
\centering
\includegraphics[width=0.45\textwidth,trim=0cm 0cm 0cm 0cm,clip]{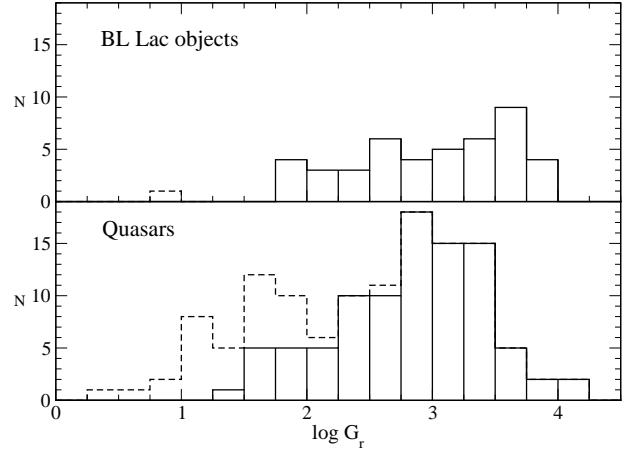}
\caption{\label{rg_histograms} Distribution of \gr to radio luminosity
  ratio for BL Lacs (top panel) and quasars (bottom panel). Upper
  limit values for AGNs with no \fermi 1LAC catalog associations are
  indicated by the dashed lines.  }
\end{figure}

None of the radio galaxies are significantly \grd loud, with ratios
all below 65.  The quasars and BL Lacs have significantly different
$G_r$ distributions (Figure~\ref{rg_histograms}), with the former
peaking at $G_r \simeq 10^3$ and the latter peaking above $10^{3.5}$.
There is a substantial population of quasars with $G_r$ values below
100, while all of the BL Lacs (with the exception of J0006$-$0623) have
\Fermi associations  and $G_r > 60$.  The Peto and Peto
modification of Gehan's Wilcoxon two-sample test for censored data
rejects the null hypothesis that the quasar and BL Lac $G_r$ values
come from the same parent population at the 99.99\% confidence level.

\begin{figure}[t]
%\epsscale{.85}
%\figurenum{1}
%\plotone{rg_vs_sedpeak.ps}
\centering
\includegraphics[width=0.5\textwidth,trim=0cm 0cm 0cm 1.7cm,clip]{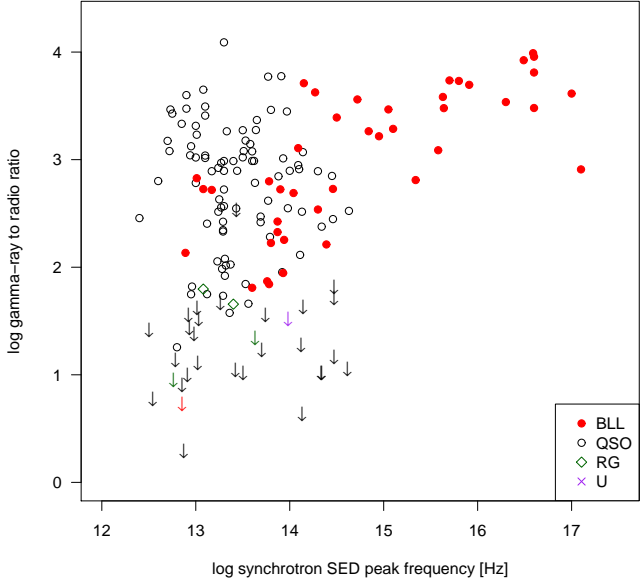}
\caption{\label{rg_vs_sedpeak}  \gr to radio luminosity
  ratio $G_r$ versus synchrotron SED peak frequency. The red filled circles represent BL Lac
  objects, the open circles quasars, the green diamonds radio galaxies,
  and the purple crosses optically unidentified objects. The arrows denote
  upper-limits. The BL Lac objects
  show a linear trend of increasing \gr loudness with
  SED peak frequency, while no trend exists for the quasars. }
\end{figure}

\begin{figure}
%\epsscale{.85}
%\figurenum{1}
%\plotone{sed_peak_vs_fluxradio.ps}
\centering
\includegraphics[width=0.5\textwidth,trim=0cm 0cm 0cm 1.7cm,clip]{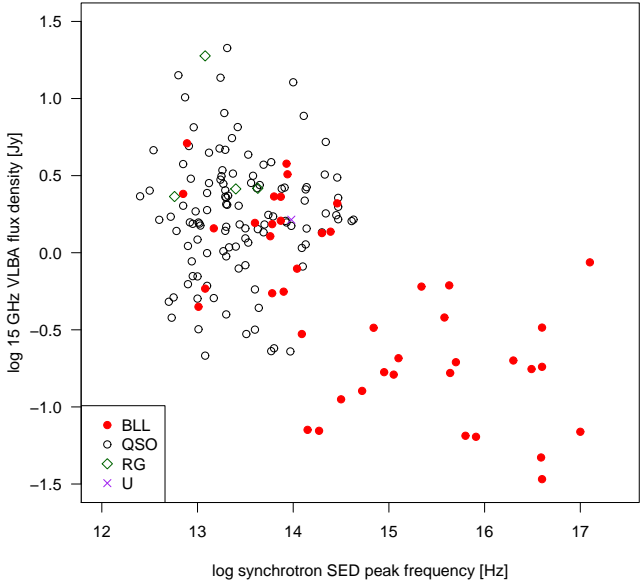}
\caption{\label{sed_peak_vs_fluxradio} 15 GHz VLBA flux density versus synchrotron SED peak frequency.  The red filled circles represent BL Lac
  objects, the open circles quasars, the green diamonds radio galaxies,
  and the purple crosses optically unidentified objects. }
\end{figure}

These differences are reflected in Figure~\ref{rg_vs_sedpeak}, which
shows \gr loudness plotted against the synchrotron SED peak frequency.
The BL Lacs show a roughly linear correlation of the form $\log{G_r} =
(0.40\pm 0.06)\log{\nu_s} - 2.9 \pm 0.9$, with a scatter of 0.5 dex,
while the quasars show no trend.  It is apparent that the BL Lacs have
a higher mean \gr loudness value because many of them
have synchrotron peaks above $\sim 10^{15}$ Hz.  Since the fixed radio
bandpass is always located below the synchrotron peak, if we compare
two BL Lacs with identical SED shapes but different synchrotron peak
locations, the high synchrotron peaked BL Lac will have a lower radio
flux density, and thus a higher \gr loudness value.
Figure~\ref{sed_peak_vs_fluxradio} shows this broad trend for the BL
Lacs, with the HSP jets having generally lower radio flux densities
than the LSPs.

\begin{figure}[t]
%\epsscale{.85}
%\figurenum{1}
%\plotone{rg_vs_energyindex.ps}
\centering
\includegraphics[width=0.5\textwidth,trim=0cm 0cm 0cm 1.1cm,clip]{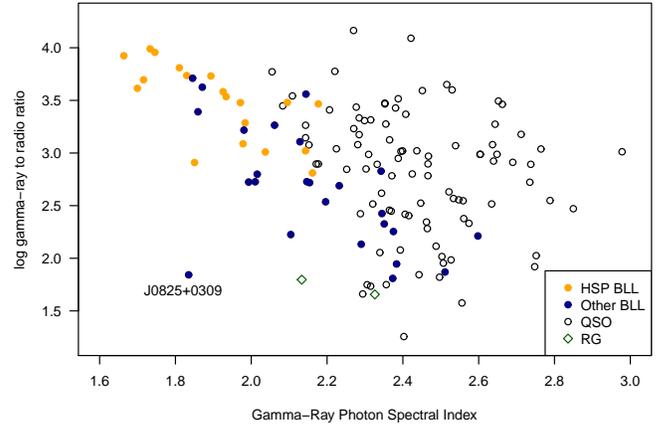}
\caption{\label{rg_vs_energyindex} Plot of \gr to radio luminosity
  ratio $G_r$ versus \gr photon spectral index. The filled circles
  represent BL Lac objects, with the high synchrotron peaked ones in
  orange and others in blue. The open circles represent quasars, and
  the green diamonds radio galaxies. The BL Lac objects show a
  log-linear trend of decreasing \gr loudness with photon spectral
  index, while no trend exists for quasars. }
\end{figure}

A similar spectral index effect also occurs as the high energy SED
peak moves in tandem through the \fermi LAT band as the synchrotron
peak frequency increases. This is manifested in the strong correlation
seen between the \gr photon spectral index $\alpha_G$ and synchrotron
peak frequency for the 1FGL blazars, as described by \cite{1LAC}. In
Figure~\ref{rg_vs_energyindex} we plot \gr loudness against photon
spectral index. Again we see a good (even tighter) linear correlation
for the BL Lac objects, and no trend for the quasars. A regression fit
to the BL Lacs, omitting the outlier source J0825+0309, gives
$\log{G_r} = (-2.3 \pm 0.2)\alpha_G + 7.8 \pm 0.5$, with a scatter of
0.3 dex.


The continuous trend from LSP to HSP BL Lacs in
Figures~\ref{rg_vs_sedpeak} and \ref{rg_vs_energyindex} is noteworthy, since it
implies a relatively narrow intrinsic range of variation in the SED
shapes of the brightest BL Lac objects. Broadly speaking, there are three
aspects of  a SED that can affect its measured \gr loudness
parameter. These are the relative positions of the synchrotron and
high energy peaks with respect to the fixed \gr and radio bands, their
relative luminosities (often referred to as the Compton dominance),
and the width and shape of each peak. If we take the simplest case
of both peaks having equal luminosity and identical parabolic forms
in $\nu F_\nu$ --$\nu$ space, then we would expect to have
\begin{equation}
\log{G_r} = C_1 \left(\log{ \nu_h \over \nu_\gamma}\right)^2 - C_2\left(\log{
    \nu_s \over \nu_r}\right)^2,
\end{equation}
where $\nu_\gamma$ and $\nu_r$ are the frequencies of the LAT \gr and
VLBA radio bands, $\nu_h$ and $\nu_s$ are the
frequencies of the high energy and synchrotron peaks, and $C_1$ and
$C_2$ are parameters that determine their respective widths.

If both SED peaks have identical parabolic shape ($C_1$ = $C_2$) and
the entire SED is then shifted to a higher frequency, such that the peak
separation $\log{\nu_h}-\log{\nu_s} = C_3 $ remains constant, then we would
expect to find a linear relation of the form $\log{G_r} = a \log{\nu_s}$, with slope
\begin{equation}
a = 2 C_1 \left(\log{\nu_r \over \nu_\gamma} + C_3\right).
\end{equation}

From compilations of observed blazar SEDs (e.g.,
\citealt{abdo_sed,Chang_thesis}) we know that in actuality, SED peak
shapes are only approximately parabolic, and that there exists a range
of $C$ parameter values among the population.  These factors would
tend to distort any trend between $\log{\nu_s}$ and $\log{G_r}$ from
the simple linear one described here.  Furthermore, an intrinsic range
of Compton dominance parameter would likely destroy any linear
relation completely.  The fact that we see a scatter of only 0.5 dex
for the BL Lac objects therefore implies that the SEDs of the
brightest AGNs of this class must have relatively similar shapes, at
least much more so than the quasars, which show no $\nu_s$-$G_r$
correlation. Our results are corroborated by a recent study of the
1LAC by \cite{2011arXiv1106.5172G}, who defined a ''Compton
efficiency'' parameter as the ratio of the high-energy (inverse
Compton) SED peak luminosity to 8 GHz radio VLA core luminosity. They found
a similar trend of higher Compton efficiency with increasing
synchrotron peak frequency for BL Lac objects, but no trend for FSRQs.

So far in this discussion we have omitted the possible effects of
relativistic beaming on the SED. For the simple case of the same
Doppler factor in both the radio and \grd emitting regions, the entire
SED should be blue-shifted by the Doppler factor, and the apparent
luminosity of both peaks will be increased by Doppler boosting. Models
which attribute the high energy peak to inverse Compton scattering of
external seed photons by relativistic electrons in the jet predict a
higher Doppler boost in $\gamma$-rays, because of the additional
Lorentz transformation between the seed photon and jet rest frames
\citep{Dermer1995}.  In this case, when considering a jet at smaller
viewing angle, the resulting increase in Doppler factor boosts the
luminosity of the high energy peak to a level much higher than the
synchrotron peak, thereby increasing the observed Compton dominance
and \gr loudness. If the seed photons are internal to the jet, for a
single-zone synchrotron self-Compton (SSC) model relatively equal
boosting is expected in both regimes; thus $G_r$ in this case is much
less sensitive to Doppler boosting. The fairly good linear
$G_r$-$\nu_s$ correlation for the BL Lacs therefore favors the SSC
process as the dominant emission mechanism in this class of blazars.
This is in general agreement with the conclusions of recent studies
which have modeled the SEDs of {\it Fermi}-detected blazars with
detailed synchrotron and inverse-Compton emission models
\citep{abdo_sed}. It should be possible to investigate this issue in
much greater depth when more detailed information on the SED
parameters of our full sample can be obtained.

% Check whether trends with T_b could be due to just a radio flux
% versus flux correlation.

\subsection{Parsec-Scale Radio Jet Properties}
\subsubsection{Core Brightness Temperature}

Nearly all of the AGNs in our sample have a parsec-scale radio jet
morphology that is dominated by a bright, flat-spectrum core, which is
often unresolved or barely resolved in our mas-scale VLBA images. At our
observing frequency of 15 GHz, this core typically represents the
region where the jet becomes optically thick, with the true jet nozzle
being located upstream \citep{2011arXiv1103.6032S}. The brightness temperature of the core
component in our VLBA images measures the compactness of the radio jet
emission, and has been previously shown to be correlated with
indicators of relativistic beaming, such as superluminal apparent
speed \citep{Homan_etal06} and radio flux density variability
\citep{Tingay01, Hovatta09}.

\begin{figure}
%\epsscale{.85}
%\figurenum{1}
%\plotone{tb_vs_sedpeak.ps}
\centering
\includegraphics[width=0.5\textwidth,trim=0cm 0cm 0cm 1.7cm,clip]{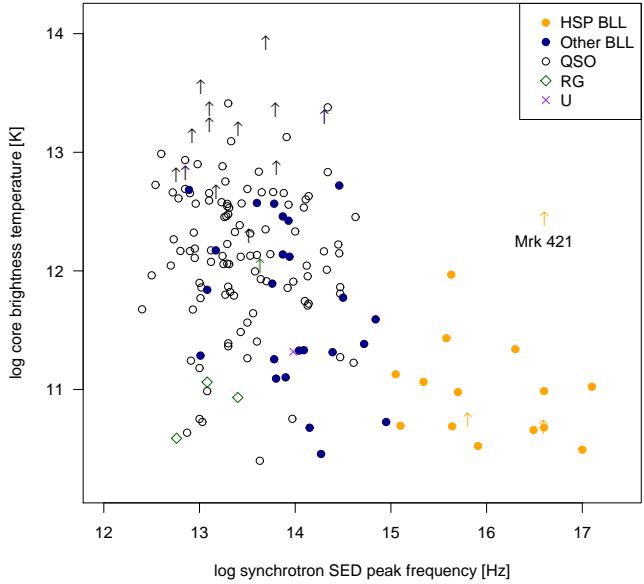}
\caption{\label{tb_vs_sedpeak} Radio core brightness temperature at 15
  GHz versus synchrotron SED peak frequency. The filled circles
  represent BL Lac objects, with the high synchrotron peaked ones in
  orange and others in blue.  The open circles represent quasars, the
  green diamonds radio galaxies, and the purple crosses
  optically unidentified objects. The arrows denote lower limits. The
  radio cores of the high-synchrotron SED peak BL Lac objects tend to
  be less compact than the other AGNs in our sample. J1104+3812 (Mrk
  421) is the only high-synchrotron peaked blazar in the sample with a
  high core brightness temperature. Not plotted is the unusually
  low-brightness temperature quasar J0957+5522 (4C +55.17) at $\nu_s =
  10^{13.77}$ Hz, $T_b = 10^{8.46}$ K.  }
\end{figure}

In Figure~\ref{tb_vs_sedpeak} we plot core brightness temperature
against synchrotron SED peak frequency. The main visible trend is that
the HSP BL Lac radio cores tend to be less compact than those of the
other AGNs in the sample. We discuss the possible ramifications of this
trend on beaming and jet velocity stratification models for
HSP BL Lacs in \S~\ref{HSP}.

\subsubsection{\label{opangles}Apparent Jet Opening Angles}

In a previous study of the MOJAVE sample using the initial 3 months of
\fermi data, \cite{Pushkarev2010} found a tendency for the \grd
detected blazars to have wider apparent opening angles than the
non-detected ones.  Since the calculated intrinsic opening angles of
the two groups were similar, they concluded that the \grd detected
jets were viewed more closely to the line of sight.

\begin{figure}
%\epsscale{.85}
%\figurenum{1}
%\plotone{rg_vs_opangle.ps}
\centering
\includegraphics[width=0.5\textwidth,trim=0cm 0cm 0cm 1.1cm,clip]{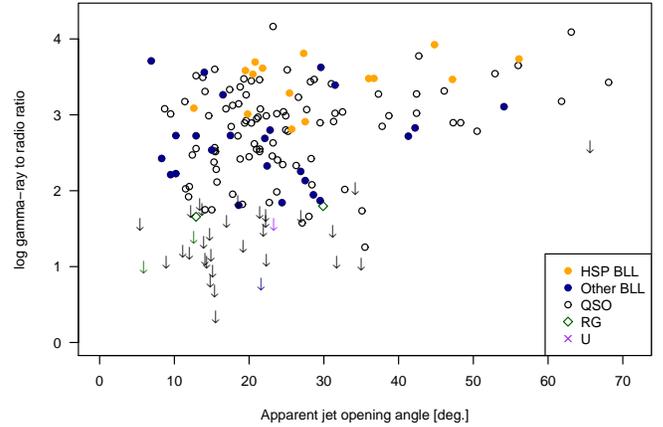}
\caption{\label{rg_vs_opangle} \gr to radio luminosity ratio versus
  apparent jet opening angle.  The filled circles represent BL Lac
  objects, with the high synchrotron peaked ones in orange and others
  in blue. The open circles represent quasars, the green diamonds
  radio galaxies, and the purple crosses optically unidentified
  objects. The AGNs with wide apparent opening angles tend to have high \gr
  loudness values. }
\end{figure}

We have analyzed the apparent jet opening angles of our sample, and
find that they range from 5 to 68 degrees, with a mean of 24 degrees.
There is an extended tail to the distribution, with 19 jets having
opening angles greater than 40 degrees.  With the exception of the
quasar J0654+4514, none of the high opening angle jets are highly
variable (radio modulation indices all less than 0.26). We find no
statistically discernible differences in the opening angle
distributions of the different optical or SED classes.  We do find a
correlation between \gr loudness and apparent opening angle, however
the relationship is non-linear (Fig.~\ref{rg_vs_opangle}). All of the
AGNs in the high-opening angle tail ($>40$ deg.) of the distribution
are significantly \grd loud ($G_r > 100$). The apparent opening angle
of a jet is related to the viewing angle and intrinsic opening angle,
with smaller intrinsic angles expected for high Lorentz factor jets
based on hydrodynamical considerations \citep{2005AJ....130.1418J}.
The high opening angle jets in our sample are a mixture of BL Lac
objects and FSRQ, with a range of synchrotron SED peak frequencies.
With jet kinematic information on the sample from the MOJAVE program
it will be possible to investigate whether these particular jets are
viewed unusually close to the line of sight, or have atypically large
intrinsic opening angles.

\subsubsection{Radio Core Polarization Vectors}

We compared the direction of the linear polarization vector at the
radio core position to the mean jet position angle for our sources, as
described in \S~\ref{vlba_data}.  In some sources such as PKS
1502$+$106 \citep{Abdo_1502+106} and PMN J0948$+$0022
\citep{Foschini2011}, changes in the core polarization angle have been
seen to occur in conjunction with \gr flaring events, suggesting a
close connection between the radio and \gr emission regions. We do not
find any correlations between the core polarization vector offset and
any \gr or SED properties for our sample. However, since the linear
polarization vector angles tend to be highly variable in blazars
\citep{2005AJ....130.1418J}, a more detailed analysis would require
truly simultaneous VLBA-\fermi measurements rather than the average
\gr data that we use in our current study. Another possible reason for
the lack of correlations is Faraday effects in the cores, which
can rotate the observed polarization vectors. We are currently
completing a VLBA rotation measure analysis of the original MOJAVE
radio-selected sample \citep{2011arXiv1108.1514H} to further investigate
this effect.

\subsubsection{Radio Core Polarization Level}

\begin{figure}[t]
%\epsscale{.85}
%\figurenum{1}
%\plotone{mcore_vs_sedpeak.ps}
\centering
\includegraphics[width=0.5\textwidth,trim=0cm 0cm 0cm 1.1cm,clip]{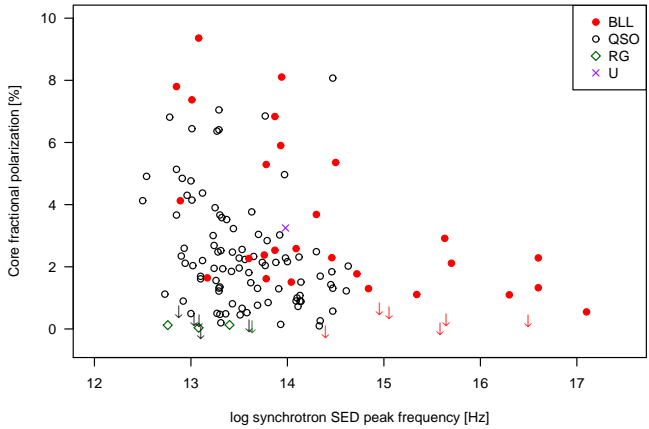}
\caption{\label{mcore_vs_sedpeak} Linear fractional
  polarization level of VLBA radio core at 15 GHz versus
  synchrotron SED peak frequency. The red filled circles represent BL Lac
  objects, the open circles quasars, the green diamonds radio galaxies,
  and the purple crosses optically unidentified objects. The arrows denote
  upper-limits.  }
\end{figure}

In Figure \ref{mcore_vs_sedpeak} we plot the linear fractional
polarization level of the VLBA core at the reference epoch versus SED
peak frequency. In general the cores of the jets are weakly polarized
($< 4$ \%), with increasing fractional polarization levels seen
downstream \citep{MOJAVE_I}.  There are no appreciable differences in
the BL Lac and FSRQ core polarization distributions, however, as we
discuss in section \ref{HSP}, the high synchrotron peak BL Lacs tend
to have low core polarization levels. We find no trend between 15 GHz
radio core polarization and \gr loudness, although in the VIPS 5 GHz
VLBA survey \citep{2011ApJ...726...16L} detected core polarization more
frequently in LAT-detected AGNs than in the non-LAT ones. We note that
the core polarizations tend to vary over time in these jets, which can
complicate such analyses. Indeed, in a preliminary investigation of
the original MOJAVE radio flux-limited sample, we found the
LAT-detected AGNs tended to have higher median fractional core
polarization levels during the first three months of the Fermi
mission, as compared to their historical average level
\citep{2010IJMPD..19..943H}. A more complete polarization analysis of
our full multi-epoch MOJAVE VLBA dataset will be presented in a
forthcoming study.

\subsubsection{Radio Variability}
\begin{figure}[t]
%\epsscale{.85}
%\figurenum{1}
%\plotone{mi_vs_sedpeak.ps}
\centering
\includegraphics[width=0.5\textwidth,trim=0cm 0cm 0cm 1.7cm,clip]{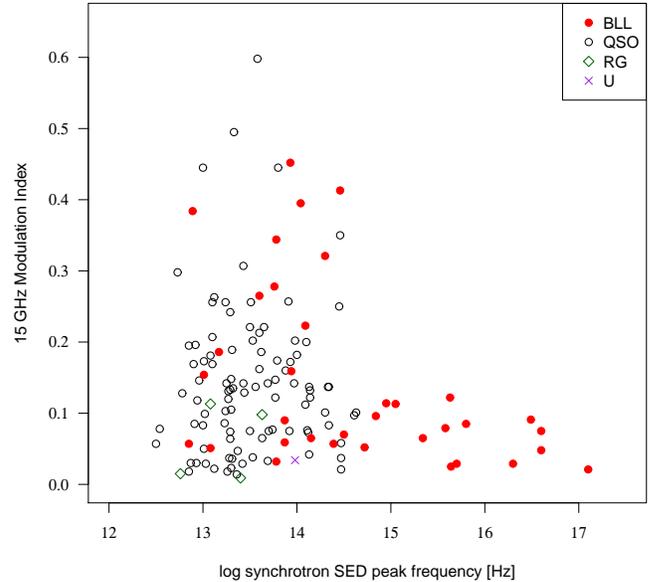}
\caption{\label{mi_vs_sedpeak} Radio modulation index at 15 GHz versus
  synchrotron SED peak frequency. The filled circles represent BL Lac
  objects, the open circles quasars, the plus symbols radio galaxies,
  and the crosses optically unidentified objects. None of the blazars
  with synchrotron SED peaks  above $10^{15}$ Hz show high amplitude radio variability.}
\end{figure}

The hallmark flux variability seen in blazars is believed to be
closely related to Doppler beaming \citep{AAH92, LV99,Hovatta09} since
it can significantly heighten the magnitudes of flaring events and
shorten their apparent timescales (e.g., \citealt{Lister01}).  AGN
jets have also been found to be in a more active radio state within several
months from LAT-detection of their strong \gr emission
\citep{MF2, 2010ApJ...722L...7P}. The AGNs in our
sample are indeed highly variable, with 51 of 144 sources having
standard deviations greater than 15\% of their mean flux density level
over an 11 month period. In their full sample of over 1000 sources,
\cite{Richards11} found the FSRQs to have significantly higher
variability amplitudes than the BL Lacs. We do not see this
distinction in our sample, however, most likely because ours contains
a smaller proportion of high-synchrotron peaked BL Lacs. The latter
tend to have moderately low radio modulation indexes, as seen in
Figure~\ref{mi_vs_sedpeak}.

\subsection{\label{HSP}High Synchrotron Peaked AGN Jets and the BL Lac Blazar Class}

Previous studies of the full 1LAC catalog by the LAT team
\citep{1LAC,abdo_sed} have established that HSP BL Lacs have
fundamentally different \gr properties than the \grd loud
FSRQs. In our study we have found that the HSP BL Lacs
are characterized by high \gr to radio luminosity ratios and lower
than average radio core compactness. Given these differences seen in
both the radio and \gr regimes, a fundamental question remains as to
whether the lower synchrotron-peaked BL Lacs also form a jet population
distinct from the FSRQs. The continuity of the trend between SED peak
frequency and \gr loudness (Figure~\ref{rg_vs_sedpeak}) would suggest
that their SED shapes are similar to the HSP BL Lacs, and thus they
should be unified with them. They are also more similar to the HSPs in
terms of their radio luminosity, as compared to the generally more
luminous FSRQs (Figure~\ref{lgam_vs_lradio}).

If we directly compare the radio properties of the LSP BL Lacs and LSP
FRSQs in our sample, we find that the LSP BL Lacs have higher mean
fractional linear polarization (4.8 $\pm$ 2.8 versus 2.5 $\pm$ 1.7;
t=3.02, p=0.992 according to a Welch Two Sample t-test), although both
classes span roughly the same range of extreme values
(Fig.~\ref{mcore_vs_sedpeak}). It is possible that beam depolarization
effects may lower the mean value for the FSRQs, since they are
typically at higher redshift than the BL Lacs and are thus imaged with
poorer spatial resolution.

However, if we compare the LSP and HSP BL Lacs, which have similar
redshift ranges, we find the latter have consistently low core
polarization and modulation indices, as well as lower than average
radio core brightness temperatures. Since high radio variability, core
polarization, and brightness temperature are generally associated with
high Doppler boosted-jets (e.g., \citealt{Lister43GHZPR, Tingay01,
  Hovatta09}), the trends we find in our sample support the following
scenario for the brightest \gr and radio blazars in the sky. Because
of their higher intrinsic \gr loudness ratios and low redshifts, the
HSP BL Lacs do not need to be as highly beamed to enter into
flux-limited \gr and radio samples, thus they tend to have lower
Doppler boosting factors than other blazar classes. The LSP BL Lacs
are less intrinsically luminous than the FSRQs, but their moderately
high intrinsic \gr loudness ratios and Doppler boosting factors
combine to give them apparent \gr and radio luminosities comparable to
the fainter end of the FSRQ distribution.

The above scenario is supported by the previous results of
\cite{Nieppola2008}, who found a general trend of decreasing
Doppler factor with increasing synchrotron SED peak frequency. Their
sample only included blazars of the LSP and ISP class, however.  A
potential test can be made with parsec-scale superluminal motion
measurements, which set an upper limit on the viewing angle and a
lower limit on the bulk jet Lorentz factor (see e.g.,
\citealt{UP95}). One of the main unresolved problems for HSP BL Lacs
has been the relatively slow apparent jet speeds detected for these
objects \citep{Piner2004, Piner2010}, despite the need for large
Doppler factors to account for rapid variability seen in $\gamma$-rays
and to accurately model their SEDs (see, e.g.,
\citealt{HenriSauge2006}). Several models have been put forward to
address this ``Doppler factor crisis'', including decelerating flows
\citep{2003ApJ...594L..27G}, and stratified spine-sheath models
\citep{Celotti2001,2008MNRAS.385L..98T}, in which the $\gamma$-rays
originate in a high-velocity jet spine, while the radio emission (and
moving blobs) are associated with a lower-Lorentz factor sheath. In
this manner the radio and \gr emission can have independent Doppler
factors. As we discussed in \S~\ref{gamrayloudness}, however,
uncorrelated beaming factors for the synchrotron and high-energy peaks
would likely destroy any linear relation between SED synchrotron peak
and \gr loudness, in contrast to what we see for the BL Lacs in our survey 
(Figure~\ref{rg_vs_sedpeak}). A more recent model put forward by
\cite{Lyutikov2010} involving non-steady magnetized outflows suggests
the existence of different, yet correlated Doppler factors for the two
SED peak regions, which can potentially preserve the $G_r$ versus
synchrotron SED peak relation.  With the MOJAVE program we are
currently obtaining multi-epoch VLBA measurements of all the \grd
selected radio jets in our sample, which will allow us to further
investigate the connections between synchrotron SED peak frequency,
apparent jet speed, jet opening angle, and Doppler factor in the
brightest blazars.

% are the same Bl Lacs that are high variab the ones that are high 
%core frac pol? --- No

\section{SUMMARY\label{summary}}

We have obtained single-epoch 15 GHz MOJAVE program VLBA
images\footnote{MOJAVE data archive:
  http://www.physics.purdue.edu/MOJAVE} of two complete flux-limited
samples of blazars north of declination $-30^\circ$ and $>10^\circ$
from the Galactic plane. The first sample consists of the brightest
sources associated with AGNs that were detected by the \fermi LAT
instrument during its first 11 months of operations. The second sample
contains all radio-loud AGNs known to have exceeded a 15 GHz VLBA flux
density of 1.5 Jy during the same time period. There are 116 AGNs in
the \grd selected sample and 105 AGNs in the matching radio one, with
48 AGNs in common to both samples. By covering two complete regions of
the \grd radio flux plane, we are able to sample the widest possible
range of \gr loudness in bright blazars. Our results can be compared
with those of \cite{2011ApJ...726...16L}, who studied a larger, radio
flux-limited sample of fainter AGNs with 5 GHz VLBA data obtained
several years before the launch of \fermi. We summarize our major
findings as follows:

1. There is a 28\% overlap in our samples of the brightest \grd and
radio-selected AGNs in the northern sky. We find no major differences
in the sample redshift distributions, with the exception of a small
number of high-synchrotron peaked (HSP) BL Lac objects which appear
only in the \grd selected sample.  At the highest flux levels
therefore, \grd and radio-selected blazars are essentially selected
from the same general parent population.

2. We have tabulated a \gr loudness parameter ($G_r$) for all of the
sources in our sample, including upper limits for the non-LAT
associated sources in our radio-selected sample. The non-censored
$G_r$ values span nearly four orders of magnitude, reflecting the wide
range of SED parameters in the bright blazar population. However,
within the BL Lac optical class, we find a linear relation (scatter =
0.5 dex) between synchrotron SED peak frequency $\nu_s$ and $G_r$,
with the HSP BL Lacs being more \grd loud. Such a linear relationship
is expected if the overall range of SED shape is relatively narrow
within the BL Lac population, as the radio flux density will drop and
\gr flux will increase as the SED is successively shifted to higher
frequencies, thereby increasing the \gr loudness. A further
consequence of the observed correlation is that the amount of
Doppler boosting must be correlated in the radio and \gr emission regimes
of BL Lac jets. The external seed-photon inverse-Compton model is not
favored for the \gr emission of the BL Lacs in our sample, since it
predicts higher boosting of the $\gamma$-rays with respect to the
radio. Any range of Doppler factor within the BL Lac population would
therefore destroy any expected linear correlation between $G_r$ and $\nu_s$ in
the external Compton scenario.

3. In terms of their 15 GHz radio properties, the HSP BL Lac objects
in our sample are distinguished by their lower than average radio core
brightness temperatures. None of them display large radio modulation
indices or high linear core polarization levels. Given the known
association of such properties with relativistic beaming, we suggest
that the HSP BL Lacs have generally lower Doppler factors than the
lower-synchrotron peaked BL Lacs or FSRQs in our sample. They are able to meet
our flux-selection criteria primarily because of their
high intrinsic \gr ratios and low redshifts. The continuity of the
observed $G_r$ - $\nu_s$ relation suggests that the high- and
low-synchrotron peaked BL Lacs are part of the same parent
population. The latter have moderate intrinsic \gr loudness ratios and
Doppler boosting factors which combine to give them apparent \gr and
radio luminosities that are comparable to the fainter end of the
FSRQ distribution. 

4. We confirm the results of a previous analysis by
\cite{Pushkarev2010}, who found that \fermi-associated AGNs tend to
have wider apparent jet opening angles. Using a larger (11 month
versus 3 month) \fermi dataset on our more comprehensive blazar sample,
we find that all of the highest opening angle jets ($> 40^\circ$) in
our sample are significantly \gr loud. 

The MOJAVE program is continuing to investigate these issues by
obtaining multi-epoch VLBA measurements of all the \grd selected radio
jets in our sample.  Together with more complete SED information,
light curves and deep EVLA images we aim to gain a fuller
understanding of the connections between synchrotron SED peak
frequency, Compton dominance, apparent jet speed, and Doppler factor
in the brightest blazars.

\acknowledgments
We thank M.~H.~Cohen and D. C. Homan for helpful comments on the manuscript. 

C. S. Chang was a former member of the International Max Planck Research  
School for Astronomy and Astrophysics.  C. S. Chang acknowledges  
support by the EU Framework 6 Marie Curie Early Stage Training  
programme under contract number MEST-CT-2005-19669 ''Estrela''.

Y.~Y.~Kovalev was supported in part by the return fellowship of the Alexander 
von Humboldt Foundation and the Russian Foundation for Basic Research 
(RFBR) grants 08-02-00545 and 11-02-00368.

E. Ros acknowledges partial support by the Spanish MICINN through grant AYA2009-13036-C02-02.

Work at UMRAO was made possible by grants from the NSF and NASA and by
support from the University of Michigan.

The \textit{Fermi} LAT Collaboration acknowledges generous ongoing
support from a number of agencies and institutes that have supported
both the development and the operation of the LAT as well as
scientific data analysis.  These include the National Aeronautics and
Space Administration and the Department of Energy in the United
States, the Commissariat \`a l'Energie Atomique and the Centre
National de la Recherche Scientifique / Institut National de Physique
Nucl\'eaire et de Physique des Particules in France, the Agenzia
Spaziale Italiana and the Istituto Nazionale di Fisica Nucleare in
Italy, the Ministry of Education, Culture, Sports, Science and
Technology (MEXT), High Energy Accelerator Research Organization (KEK)
and Japan Aerospace Exploration Agency (JAXA) in Japan, and the
K.~A.~Wallenberg Foundation, the Swedish Research Council and the
Swedish National Space Board in Sweden.

Additional support for science analysis during the operations phase is
gratefully acknowledged from the Istituto Nazionale di Astrofisica in
Italy and the Centre National d'\'Etudes Spatiales in France.

The MOJAVE project is supported under National Science Foundation grant
AST-0807860 and NASA \Fermi grant NNX08AV67G.

This research has made use of the NASA/IPAC Extragalactic Database
(NED) which is operated by the Jet Propulsion Laboratory, California
Institute of Technology, under contract with the National Aeronautics
and Space Administration.

This work made use of the Swinburne University of Technology software 
correlator \citep{2011PASP..123..275D}, developed as part of the Australian Major National Research  Facilities Programme and operated under licence. 

The National Radio Astronomy Observatory is a facility of the National Science Foundation operated under cooperative agreement by Associated Universities, Inc.

\facility[NRAO(VLBA)]{The VLBA is a facility of the National Science
Foundation operated by the National Radio Astronomy Observatory under
cooperative agreement with Associated Universities, Inc.}

{\it Facilities:} \facility{VLBA, \Fermi(LAT), UMRAO, OVRO}.

%\appendix
\begin{appendix}
\section{Redshift Information}

We summarize the current status of those blazars in
our sample which either do not have a spectroscopic redshift listed in
Table 1, or have uncertain or conflicting reported redshift values in the literature.

\paragraph{J0050$-$0929:} The NED value of $z=0.634$ from \cite{2001AJ....122..565R}
is based on a very weak emission line and is listed by those authors
as tentative. This line was not seen by \cite{2009ApJ...704..477S} or
\cite{2006AJ....132....1S} in their optical spectra. Based on the
absence of host galaxy absorption features in the optical spectrum
\cite{2009ApJ...704..477S} find $z > 0.44$, and
\cite{2006AJ....132....1S} find $z > 0.3$. \cite{2010ApJ...712...14M}
obtain $ z > 0.27$ based on the measured optical host galaxy
magnitude.

\paragraph{J0112+2244:} Healey et al. (2008) list $z= 0.265$ based on an unpublished
spectrum. \cite{2009ApJ...704..477S} find $z > 0.24$ and
\cite{2009AJ....137..337S} get $z > 0.25$ based on the absence of host
galaxy absorption features in the optical spectrum.

\paragraph{J0120$-$2701:} The NED value of $z=0.559$ is a lower limit
from \cite{SFK93}. 

\paragraph{J0136+3908:}  We could not find any published redshift value for
this source. 

\paragraph{J0237+2848:} The NED value of $z=1.213$ attributed to
  \cite{Schmidt77} is different than the $z=1.207$ that is reported in
  that reference.

\paragraph{J0222+4302 = 3C 66A:} As discussed by \cite{FSB08} and
\cite{2005ApJ...629..108B}, the NED redshift value of $z = 0.444$
is highly unreliable. \cite{FSB08} obtained $z > 0.096$ based on
their optical spectrum.

\paragraph{J0316+0904:} We could not find any published redshift value for
this source. 

\paragraph{J0433+2905:} \cite{2010ApJ...712...14M} obtain $ z > 0.48$ based
  on the measured optical host galaxy magnitude. The origin of the $z
  = 0.97$ value listed in BZCAT \citep{2009A&A...495..691M} is
  unknown.

%not in 1FM paper:} 0454+844:} \cite{BVC08} list $z = 0.112$ based on an
%unpublished %spectrum, however,\cite{1997ApJ...489L..17S} find an absorption
%feature at z = 1.34. G. Taylor in email Jun 2011 is not so sure anymore about
% published 2008 result.

\paragraph{J0509+0541:} \cite{2010ApJ...712...14M} obtain $ z > 0.38$ based on the
measured optical host galaxy magnitude.

\paragraph{J0608$-$1520:}   \cite{Shaw2011} have obtained an optical spectrum of this source and find emission lines indicating a quasar at $z = 1.094$.

\paragraph{J0612+4122:} \cite{2010ApJ...712...14M} obtain $ z > 0.69$ based on the
measured optical host galaxy magnitude. 

\paragraph{J0630-2406:} \cite{Landt2008} list $z = 1.238$, which is a lower
  limit based on Mg II absorption lines in their unpublished optical spectrum.

\paragraph{J0654+5042:}   \cite{Shaw2011} have obtained an optical spectrum of this source and find emission lines indicating a quasar at $z = 1.253$. 

\paragraph{J0738+1742:} The NED value of $z = 0.424$ is a lower limit
  determined by \cite{2001AJ....122..565R} on the basis of absorption
  systems in the optical spectrum.

\paragraph{J0818+4222:} \cite{BVC08} list $z = 0.245$ based on an
  unpublished spectrum. The NED value of $z = 0.53$ is attributed to
  \cite{SRM05} but does not appear in that paper. An optical spectrum
  obtained by \cite{Shaw2011} shows no visible
  features. \cite{2005ApJ...635..173S} found $z > 0.75$ based on a
  lower limit to the host galaxy magnitude.

\paragraph{J1037+5711:} \cite{2000A&ARv..10...81V} classify this source as a BL
Lac object. We were unable to find any redshift value in the literature.    

\paragraph{J1248+5820:} The SDSS optical spectrum \citep{SDSS3} yields no
  reliable redshift, and an unpublished spectrum by \cite{Shaw2011} shows no spectral features.

\paragraph{J1215$-$1731:}  We were not able to find any published
redshift value for this optically unidentified source, which lies
extremely close on the sky to a bright star.

\paragraph{J1221+2813 = W Comae:} The NED redshift value of $z=0.102$ is likely
incorrect, as discussed by \cite{FSB08}. The latter authors constrain
the redshift to $z > 0.104$ using their optical spectrum.

\paragraph{J1303+2433:} \cite{GHW07} list $z=0.993$ but give no reference for the origin of this redshift.  An optical spectrum obtained by \cite{Shaw2011} yields $z > 0.769$.

\paragraph{J1427+2348:} \cite{2009ApJ...704..477S} find $z > 0.03$, while \cite{2010ApJ...712...14M} get $z > 0.23$ based on host galaxy magnitude.

\paragraph{J1516+1932:} The NED redshift value of $z=1.07$ from \cite{PS86}
  is based on tentative identifications of very faint emission lines
  in the optical spectrum of \cite{1983PASAu...5....2W}, and has not
  been subsequently confirmed.  An optical spectrum obtained by \cite{Shaw2011} shows no spectral features.

\paragraph{J1532+6129:} \cite{2009ApJ...704..477S} find $z > 0.63$ based on the absence of
host galaxy absorption features in the optical spectrum, while
\cite{2010ApJ...712...14M} obtain $ z > 0.39$ based on the measured
optical host galaxy magnitude.

\paragraph{J1555+1111:} The $z = 0.360$ redshift value from
\cite{1983BAAS...15..957M} was shown by \cite{1990PASP..102.1120F} and
\cite{1994ApJS...93..125F}  to be incorrect. Based on the absence of host galaxy
absorption features in the optical spectrum, \cite{2006AJ....132....1S}
find $z > 0.09$. \cite{2010ApJ...720..976D} analyze the Lyman absorber
properties of the far-UV spectrum of this source and find $0.395 < z < 0.58$.

\paragraph{J1719+1745:} \cite{SRM05} list $z = 0.137$ based on an unpublished
optical spectrum. \cite{2009ApJ...704..477S} obtain $z > 0.58$ based
on the measured optical host galaxy magnitude.

\paragraph{J1725+1152:} As described by \cite{2006AJ....132....1S},
the tentative NED redshift of $z= 0.018$ from \cite{1989MNRAS.240...33G}
has not been confirmed in several subsequent spectroscopic
observations. Based on the absence of host galaxy absorption features
in the optical spectrum, \cite{2006AJ....132....1S} find $z > 0.17$.

\paragraph{J1903+5540:} \cite{2010ApJ...712...14M}  obtain $ z > 0.58$ based on the
measured optical host galaxy magnitude.

\paragraph{J2236$-$1433:} \cite{2006AJ....132....1S} trace the oft-cited
erroneous NED redshift of $z = 0.325$ to a clerical error. Based on
the absence of host galaxy absorption features in the optical
spectrum, they find find $z > 0.65$. An unpublished optical spectrum
by \cite{Shaw2011} shows no spectral features.

\paragraph{J2243+2021:} \cite{2010ApJ...712...14M} obtain $ z > 0.39$ based on the
measured optical host galaxy magnitude.

\end{appendix}

\bibliographystyle{apj}
\bibliography{lister}

\LongTables
\begin{deluxetable*}{lllllccccc} 
\tablecolumns{10} 
%\tabletypesize{\scriptsize} 
\tablewidth{0pt}  
\tablecaption{\label{1gentable} General Properties of AGNs in the Combined $\gamma$-ray and Radio Samples}  
\tablehead{\colhead{J2000} & \colhead{B1950} &\colhead{1FGL Name} & \colhead {Alias} &  
\colhead{z} &  \colhead{Ref.} & \colhead{Opt.} &\colhead{SED} & \colhead{Ref.} &    \colhead{Sample} \\ 
\colhead{(1)} & \colhead{(2)} & \colhead{(3)} & \colhead{(4)} & \colhead{(5)} & 
 \colhead{(6)} & \colhead{(7)} & \colhead{(8)}  & \colhead{(9)}  & \colhead{(10)}   } 
\startdata 
J0006$-$0623 & 0003$-$066 &\n & NRAO 005& 0.3467& \cite{JRS09} & B& LSP& 1 &  R \\ 
J0017$-$0512 & 0015$-$054 & J0017.4$-$0510 & PMN J0017$-$0512& 0.226& \cite{Shaw2011} & Q& LSP& 2 &  G \\ 
J0050$-$0929 & 0048$-$097 & J0050.6$-$0928 & PKS 0048$-$09& \n& \n & B& ISP& 2 &  B \\ 
J0108+0135 & 0106+013 & J0108.6+0135 & 4C +01.02& 2.099& \cite{1995AJ....109.1498H} & Q& ISP& 1 &  B \\ 
J0112+2244 & 0109+224 & J0112.0+2247 & S2 0109+22& 0.265& \cite{2008ApJS..175...97H} & B& ISP& 1 &  G \\ 
J0112+3208 & 0110+318 & J0112.9+3207 & 4C +31.03& 0.603& \cite{WW76} & Q& LSP& 11 &  G \\ 
J0118$-$2141 & 0116$-$219 & J0118.7$-$2137 & OC -228& 1.165& \cite{WAA83} & Q& LSP& 2 &  G \\ 
J0120$-$2701 & 0118$-$272 & J0120.5$-$2700 & OC -230.4& \n& \n & B& LSP& 2 &  G \\ 
J0121+1149 & 0119+115 &\n & PKS 0119+11& 0.570& \cite{1994AAS..105..211S} & Q& LSP& 1 &  R \\ 
J0132$-$1654 & 0130$-$171 & J0132.6$-$1655 & OC -150& 1.020& \cite{WAA83} & Q& LSP& 11 &  B \\ 
J0136+3905 & 0133+388 & J0136.5+3905 & B3 0133+388& \n& \n & B& HSP& 5 &  G \\ 
J0136+4751 & 0133+476 & J0137.0+4751 & DA 55& 0.859& \cite{1996ApJS..107..541L} & Q& LSP& 1 &  B \\ 
J0145$-$2733 & 0142$-$278 & J0144.9$-$2732 & OC -270& 1.148& \cite{1999ApJS..122...29B} & Q& LSP& 2 &  G \\ 
J0205+3212 & 0202+319 & J0205.3+3217 & B2 0202+31& 1.466& \cite{1970ApJ...160L..33B} & Q& LSP& 1 &  R \\ 
J0204$-$1701 & 0202$-$172 & J0205.0$-$1702 & PKS 0202$-$17& 1.739& \cite{JRS09} & Q& LSP& 2 &  R \\ 
J0217+7349 & 0212+735 & J0217.8+7353 & S5 0212+73& 2.367& \cite{1996ApJS..107..541L} & Q& LSP& 1 &  R \\ 
J0217+0144 & 0215+015 & J0217.9+0144 & OD 026& 1.715& \cite{1988AA...192....1B} & Q& LSP& 1 &  B \\ 
J0222+4302 & 0219+428 & J0222.6+4302 & 3C 66A& \n& \n & B& HSP& 5 &  G \\ 
J0231+1322 & 0229+131 &\n & 4C +13.14& 2.059& \cite{1994ApJ...436..678O} & Q& LSP& 6 &  R \\ 
J0237+2848 & 0234+285 & J0237.9+2848 & 4C 28.07& 1.206& \cite{Shaw2011} & Q& LSP& 1 &  B \\ 
J0238+1636 & 0235+164 & J0238.6+1637 & AO 0235+164& 0.940& \cite{1987ApJ...318..577C} & Q& LSP& 1 &  B \\ 
J0252$-$2219 & 0250$-$225 & J0252.8$-$2219 & OD -283& 1.419& \cite{Shaw2011} & Q& LSP& 11 &  G \\ 
J0303$-$2407 & 0301$-$243 & J0303.5$-$2406 & PKS 0301$-$243& 0.260& \cite{FU00} & B& HSP& 2 &  G \\ 
J0316+0904 & 0313+085 & J0316.1+0904 & BZB J0316+0904& \n& \n & B& HSP& 5 &  G \\ 
J0319+4130 & 0316+413 & J0319.7+4130 & 3C 84& 0.0176& \cite{1992ApJS...83...29S} & G& LSP& 4 &  B \\ 
J0339$-$0146 & 0336$-$019 & J0339.2$-$0143 & CTA 26& 0.852& \cite{1978ApJS...36..317W} & Q& LSP& 1 &  R \\ 
J0349$-$2102 & 0347$-$211 & J0349.9$-$2104 & OE -280& 2.944& \cite{EYH01} & Q& LSP& 2 &  G \\ 
J0403+2600 & 0400+258 &\n & CTD 026& 2.109& \cite{Schmidt77} & Q& \n& \n &  R \\ 
J0423$-$0120 & 0420$-$014 & J0423.2$-$0118 & PKS 0420$-$01& 0.9161& \cite{JRS09} & Q& LSP& 1 &  B \\ 
J0433+0521 & 0430+052 &\n & 3C 120& 0.033& \cite{1988PASP..100.1423M} & G& LSP& 1 &  R \\ 
J0433+2905 & 0430+289 & J0433.5+2905 & BZB J0433+2905& \n& \n & B& ISP& 5 &  G \\ 
J0442$-$0017 & 0440$-$003 & J0442.7$-$0019 & NRAO 190& 0.844& \cite{Schmidt77} & Q& LSP& 6 &  G \\ 
J0453$-$2807 & 0451$-$282 & J0453.2$-$2805 & OF -285& 2.559& \cite{WAA83} & Q& LSP& 4 &  B \\ 
J0457$-$2324 & 0454$-$234 & J0457.0$-$2325 & PKS 0454$-$234& 1.003& \cite{1989AAS...80..103S} & Q& LSP& 2 &  B \\ 
J0507+6737 & 0502+675 & J0507.9+6738 & 1ES 0502+675& 0.416& \cite{2002MNRAS.336..945L} & B& HSP& 2 &  G \\ 
J0509+0541 & 0506+056 & J0509.3+0540 & TXS 0506+056& \n& \n & B& HSP& 5 &  G \\ 
J0530+1331 & 0528+134 & J0531.0+1331 & PKS 0528+134& 2.070& \cite{1993ApJ...409..134H} & Q& LSP& 1 &  B \\ 
J0532+0732 & 0529+075 & J0532.9+0733 & OG 050& 1.254& \cite{SRM05} & Q& LSP& 1 &  B \\ 
J0608$-$1520 & 0605$-$153 & J0608.0$-$1521 & PMN J0608$-$1520& 1.094& \cite{Shaw2011} & Q& LSP& 11 &  G \\ 
J0609$-$1542 & 0607$-$157 &\n & PKS 0607$-$15& 0.3226& \cite{JRS09} & Q& LSP& 1 &  R \\ 
J0612+4122 & 0609+413 & J0612.7+4120 & B3 0609+413& \n& \n & B& \n& \n &  G \\ 
J0630$-$2406 & 0628$-$240 & J0630.9$-$2406 & TXS 0628$-$240& \n& \n & B& ISP& 4 &  G \\ 
J0646+4451 & 0642+449 &\n & OH 471& 3.396& \cite{1994ApJ...436..678O} & Q& LSP& 1 &  R \\ 
J0654+4514 & 0650+453 & J0654.3+4514 & B3 0650+453& 0.928& \cite{Shaw2011} & Q& LSP& 2 &  G \\ 
J0654+5042 & 0650+507 & J0654.4+5042 & GB6 J0654+5042& 1.253& \cite{Shaw2011} & Q& LSP& 11 &  G \\ 
J0713+1935 & 0710+196 & J0714.0+1935 & WB92 0711+1940& 0.540& \cite{Shaw2011} & Q& LSP& 11 &  G \\ 
J0719+3307 & 0716+332 & J0719.3+3306 & B2 0716+33& 0.779& \cite{WBG00} & Q& LSP& 2 &  G \\ 
J0721+7120 & 0716+714 & J0721.9+7120 & S5 0716+71& 0.310& \cite{NPS08} & B& ISP& 5 &  B \\ 
J0738+1742 & 0735+178 & J0738.2+1741 & OI 158& \n& \n & B& LSP& 1 &  G \\ 
J0739+0137 & 0736+017 & J0739.1+0138 & OI 061& 0.1894& \cite{HK09} & Q& ISP& 1 &  B \\ 
J0748+2400 & 0745+241 &\n & PKS 0745+241& 0.4092& \cite{SDSS3} & Q& LSP& 4 &  R \\ 
J0750+1231 & 0748+126 & J0750.6+1235 & OI 280& 0.889& \cite{1979ApJ...232..400P} & Q& LSP& 1 &  R \\ 
J0808$-$0751 & 0805$-$077 & J0808.2$-$0750 & PKS 0805$-$07& 1.837& \cite{1988ApJ...327..561W} & Q& LSP& 4 &  B \\ 
J0818+4222 & 0814+425 & J0818.2+4222 & OJ 425& \n& \n & B& LSP& 1 &  B \\ 
J0825+0309 & 0823+033 & J0825.9+0309 & PKS 0823+033& 0.506& \cite{SFK93} & B& LSP& 1 &  R \\ 
J0830+2410 & 0827+243 & J0830.5+2407 & OJ 248& 0.942& \cite{Shaw2011} & Q& LSP& 1 &  R \\ 
J0836$-$2016 & 0834$-$201 &\n & PKS 0834$-$20& 2.752& \cite{FKW83} & Q& \n& \n &  R \\ 
J0841+7053 & 0836+710 & J0842.2+7054 & 4C +71.07& 2.218& \cite{MRR99} & Q& LSP& 1 &  R \\ 
J0854+2006 & 0851+202 & J0854.8+2006 & OJ 287& 0.306& \cite{1989AAS...80..103S} & B& LSP& 1 &  B \\ 
J0909+0121 & 0906+015 & J0909.0+0126 & 4C +01.24& 1.0256& \cite{Shaw2011} & Q& ISP& 1 &  B \\ 
J0920+4441 & 0917+449 & J0920.9+4441 & S4 0917+44& 2.189& \cite{2004SDSS2.C...0000:} & Q& LSP& 6 &  B \\ 
J0927+3902 & 0923+392 &\n & 4C +39.25& 0.695& \cite{SDSS3} & Q& LSP& 1 &  R \\ 
J0948+4039 & 0945+408 &\n & 4C +40.24& 1.249& \cite{SDSS3} & Q& LSP& 1 &  R \\ 
J0948+0022 & 0946+006 & J0949.0+0021 & PMN J0948+0022& 0.585& \cite{2004SDSS2.C...0000:} & Q& LSP& 2 &  G \\ 
J0957+5522 & 0954+556 & J0957.7+5523 & 4C +55.17& 0.8993& \cite{Shaw2011} & Q& LSP& 6 &  G \\ 
J0958+6533 & 0954+658 & J1000.1+6539 & S4 0954+65& 0.367& \cite{2001AJ....122..565R} & B& LSP& 5 &  R \\ 
J1012+2439 & 1009+245 & J1012.7+2440 & GB6 J1012+2439& 1.805& \cite{Shaw2011} & Q& \n& \n &  G \\ 
J1015+4926 & 1011+496 & J1015.1+4927 & 7C 1011+4941& 0.212& \cite{2007ApJ...667L..21A} & B& HSP& 2 &  G \\ 
J1016+0513 & 1013+054 & J1016.1+0514 & TXS 1013+054& 1.713& \cite{2004SDSS2.C...0000:} & Q& \n& \n &  G \\ 
J1037+5711 & 1034+574 & J1037.7+5711 & GB6 J1037+5711& \n& \n & B& ISP& 5 &  G \\ 
J1037$-$2934 & 1034$-$293 &\n & PKS 1034$-$293& 0.312& \cite{SF97} & Q& LSP& 10 &  R \\ 
J1038+0512 & 1036+054 &\n & PKS 1036+054& 0.473& \cite{2008ApJS..175...97H} & Q& LSP& 1 &  R \\ 
J1058+0133 & 1055+018 & J1058.4+0134 & 4C +01.28& 0.888& \cite{Shaw2011} & Q& LSP& 1 &  B \\ 
J1058+5628 & 1055+567 & J1058.6+5628 & 7C 1055+5644& 0.143& \cite{2004SDSS2.C...0000:} & B& HSP& 5 &  G \\ 
J1104+3812 & 1101+384 & J1104.4+3812 & Mrk 421& 0.0308& \cite{1975ApJ...198..261U} & B& HSP& 2 &  G \\ 
J1121$-$0553 & 1118$-$056 & J1121.5$-$0554 & PKS 1118$-$05& 1.297& \cite{DWF97} & Q& LSP& 11 &  G \\ 
J1127$-$1857 & 1124$-$186 & J1126.8$-$1854 & PKS 1124$-$186& 1.048& \cite{DWF97} & Q& ISP& 1 &  R \\ 
J1130$-$1449 & 1127$-$145 & J1130.2$-$1447 & PKS 1127$-$14& 1.184& \cite{1986MNRAS.218..331W} & Q& LSP& 2 &  B \\ 
J1159+2914 & 1156+295 & J1159.4+2914 & 4C +29.45& 0.7246& \cite{Shaw2011} & Q& ISP& 1 &  B \\ 
J1215$-$1731 & 1213$-$172 &\n & PKS 1213$-$17& \n& \n & U& LSP& 1 &  R \\ 
J1217+3007 & 1215+303 & J1217.7+3007 & ON 325& 0.130& \cite{2003ApJS..148..275A} & B& HSP& 5 &  G \\ 
J1221+3010 & 1218+304 & J1221.3+3008 & B2 1218+30& 0.1836& \cite{2007SDSS6.C...0000:} & B& HSP& 9 &  G \\ 
J1221+2813 & 1219+285 & J1221.5+2814 & W Comae& \n& \n & B& ISP& 5 &  G \\ 
J1224+2122 & 1222+216 & J1224.7+2121 & 4C +21.35& 0.434& \cite{SDSS7} & Q& LSP& 6 &  G \\ 
J1229+0203 & 1226+023 & J1229.1+0203 & 3C 273& 0.1583& \cite{1992ApJS...83...29S} & Q& LSP& 1 &  B \\ 
J1230+1223 & 1228+126 & J1230.8+1223 & M87& 0.00436& \cite{2000MNRAS.313..469S} & G& LSP& 7 &  R \\ 
J1239+0443 & 1236+049 & J1239.5+0443 & BZQ J1239+0443& 1.761& \cite{Shaw2011} & Q& LSP& 11 &  G \\ 
J1246$-$2547 & 1244$-$255 & J1246.7$-$2545 & PKS 1244$-$255& 0.633& \cite{SBB76} & Q& LSP& 2 &  G \\ 
J1248+5820 & 1246+586 & J1248.2+5820 & PG 1246+586& \n& \n & B& HSP& 5 &  G \\ 
J1256$-$0547 & 1253$-$055 & J1256.2$-$0547 & 3C 279& 0.536& \cite{1996ApJS..104...37M} & Q& LSP& 1 &  B \\ 
J1303+2433 & 1300+248 & J1303.0+2433 & VIPS 0623 & \n& \n & B& \n& \n &  G \\ 
J1310+3220 & 1308+326 & J1310.6+3222 & OP 313& 0.9973& \cite{Shaw2011} & Q& ISP& 1 &  B \\ 
J1332$-$0509 & 1329$-$049 & J1331.9$-$0506 & OP -050& 2.150& \cite{TDD90} & Q& LSP& 2 &  G \\ 
J1332$-$1256 & 1329$-$126 & J1332.6$-$1255 & PMN J1332$-$1256& 1.492& \cite{Shaw2011} & Q& \n& \n &  G \\ 
J1337$-$1257 & 1334$-$127 & J1337.7$-$1255 & PKS 1335$-$127& 0.539& \cite{1993AAS...97..483S} & Q& LSP& 1 &  B \\ 
J1344$-$1723 & 1341$-$171 & J1344.2$-$1723 & PMN J1344$-$1723& 2.506& \cite{Shaw2011} & Q& \n& \n &  G \\ 
J1427+2348 & 1424+240 & J1426.9+2347 & OQ +240& \n& \n & B& HSP& 5 &  G \\ 
J1436+6336 & 1435+638 &\n & VIPS 0792& 2.066& \cite{MRR99} & Q& LSP& 4 &  R \\ 
J1504+1029 & 1502+106 & J1504.4+1029 & OR 103& 1.8385& \cite{2007SDSS6.C...0000:} & Q& LSP& 1 &  B \\ 
J1512$-$0905 & 1510$-$089 & J1512.8$-$0906 & PKS 1510$-$08& 0.360& \cite{TDD90} & Q& LSP& 1 &  B \\ 
J1516+1932 & 1514+197 & J1516.9+1928 & PKS 1514+197& \n& \n & B& LSP& 5 &  R \\ 
J1517$-$2422 & 1514$-$241 & J1517.8$-$2423 & AP Librae& 0.049& \cite{JRS09} & B& LSP& 2 &  B \\ 
J1522+3144 & 1520+319 & J1522.1+3143 & B2 1520+31& 1.484& \cite{Shaw2011} & Q& LSP& 2 &  G \\ 
J1542+6129 & 1542+616 & J1542.9+6129 & GB6 J1542+6129& \n& \n & B& ISP& 5 &  G \\ 
J1549+0237 & 1546+027 & J1549.3+0235 & PKS 1546+027& 0.414& \cite{2004SDSS2.C...0000:} & Q& LSP& 1 &  R \\ 
J1550+0527 & 1548+056 & J1550.7+0527 & 4C +05.64& 1.417& \cite{Shaw2011} & Q& LSP& 1 &  R \\ 
J1553+1256 & 1551+130 & J1553.4+1255 & OR +186& 1.308& \cite{SDSS7} & Q& \n& \n &  G \\ 
J1555+1111 & 1553+113 & J1555.7+1111 & PG 1553+113& \n& \n & B& HSP& 5 &  G \\ 
J1613+3412 & 1611+343 & J1613.5+3411 & DA 406& 1.40& \cite{Shaw2011} & Q& LSP& 1 &  R \\ 
J1625$-$2527 & 1622$-$253 & J1625.7$-$2524 & PKS 1622$-$253& 0.786& \cite{SDM94} & Q& LSP& 2 &  B \\ 
J1635+3808 & 1633+382 & J1635.0+3808 & 4C +38.41& 1.813& \cite{Shaw2011} & Q& LSP& 1 &  B \\ 
J1638+5720 & 1637+574 &\n & OS 562& 0.751& \cite{1996ApJS..104...37M} & Q& ISP& 1 &  R \\ 
J1640+3946 & 1638+398 &\n & NRAO 512& 1.666& \cite{1989AAS...80..103S} & Q& LSP& 1 &  R \\ 
J1642+3948 & 1641+399 & J1642.5+3947 & 3C 345& 0.593& \cite{1996ApJS..104...37M} & Q& ISP& 1 &  B \\ 
J1642+6856 & 1642+690 &\n & 4C +69.21& 0.751& \cite{1996ApJS..107..541L} & Q& LSP& 6 &  R \\ 
J1653+3945 & 1652+398 & J1653.9+3945 & Mrk 501& 0.0337& \cite{SFK93} & B& HSP& 2 &  G \\ 
J1658+0741 & 1655+077 &\n & PKS 1655+077& 0.621& \cite{1986MNRAS.218..331W} & Q& LSP& 1 &  R \\ 
J1700+6830 & 1700+685 & J1700.1+6830 & TXS 1700+685& 0.301& \cite{1997MNRAS.290..380H} & Q& LSP& 4 &  G \\ 
J1719+1745 & 1717+178 & J1719.2+1745 & OT 129& 0.137& \cite{SRM05} & B& LSP& 5 &  G \\ 
J1725+1152 & 1722+119 & J1725.0+1151 & 1H 1720+117& \n& \n & B& HSP& 5 &  G \\ 
J1727+4530 & 1726+455 & J1727.3+4525 & S4 1726+45& 0.717& \cite{1997MNRAS.290..380H} & Q& LSP& 1 &  R \\ 
J1733$-$1304 & 1730$-$130 & J1733.0$-$1308 & NRAO 530& 0.902& \cite{1984PASP...96..539J} & Q& LSP& 1 &  B \\ 
J1734+3857 & 1732+389 & J1734.4+3859 & OT 355& 0.975& \cite{Shaw2011} & Q& LSP& 11 &  G \\ 
J1740+5211 & 1739+522 & J1740.0+5209 & 4C +51.37& 1.379& \cite{1984MNRAS.211..105W} & Q& LSP& 1 &  G \\ 
J1743$-$0350 & 1741$-$038 &\n & PKS 1741$-$03& 1.054& \cite{1988ApJ...327..561W} & Q& LSP& 1 &  R \\ 
J1751+0939 & 1749+096 & J1751.5+0937 & 4C +09.57& 0.322& \cite{1988AA...191L..16S} & B& LSP& 1 &  B \\ 
J1753+2848 & 1751+288 &\n & B2 1751+28& 1.118& \cite{2008ApJS..175...97H} & Q& LSP& 1 &  R \\ 
J1801+4404 & 1800+440 &\n & S4 1800+44& 0.663& \cite{1982MNRAS.200..191W} & Q& ISP& 1 &  R \\ 
J1800+7828 & 1803+784 & J1800.4+7827 & S5 1803+784& 0.6797& \cite{1996ApJS..107..541L} & B& LSP& 1 &  B \\ 
J1806+6949 & 1807+698 & J1807.0+6945 & 3C 371& 0.051& \cite{1992AAS...96..389d} & B& ISP& 1 &  B \\ 
J1824+5651 & 1823+568 & J1824.0+5651 & 4C +56.27& 0.664& \cite{Shaw2011} & B& LSP& 1 &  B \\ 
J1829+4844 & 1828+487 & J1829.8+4845 & 3C 380& 0.692& \cite{1996ApJS..107..541L} & Q& LSP& 4 &  R \\ 
J1842+6809 & 1842+681 &\n & GB6 J1842+6809& 0.472& \cite{XLR94} & Q& \n& \n &  R \\ 
J1848+3219 & 1846+322 & J1848.5+3224 & B2 1846+32A& 0.798& \cite{SRM05} & Q& LSP& 2 &  G \\ 
J1849+6705 & 1849+670 & J1849.3+6705 & S4 1849+67& 0.657& \cite{SK93} & Q& LSP& 1 &  B \\ 
J1903+5540 & 1902+556 & J1903.0+5539 & TXS 1902+556& \n& \n & B& ISP& 5 &  G \\ 
J1911$-$2006 & 1908$-$201 & J1911.2$-$2007 & PKS B1908$-$201& 1.119& \cite{HEM03} & Q& LSP& 2 &  B \\ 
J1923$-$2104 & 1920$-$211 & J1923.5$-$2104 & OV -235& 0.874& \cite{HEM03} & Q& LSP& 2 &  B \\ 
J1924$-$2914 & 1921$-$293 & J1925.2$-$2919 & PKS B1921$-$293& 0.3526& \cite{JRS09} & Q& LSP& 10 &  R \\ 
J1927+7358 & 1928+738 &\n & 4C +73.18& 0.302& \cite{1996ApJS..104...37M} & Q& ISP& 1 &  R \\ 
J1954$-$1123 & 1951$-$115 & J1954.8$-$1124 & TXS 1951$-$115& 0.683& \cite{Shaw2011} & Q& LSP& 11 &  G \\ 
J1955+5131 & 1954+513 &\n & \n& 1.223& \cite{1996ApJS..107..541L} & Q& LSP& 7 &  R \\ 
J2000$-$1748 & 1958$-$179 & J2000.9$-$1749 & PKS 1958$-$179& 0.652& \cite{1LAC} & Q& LSP& 1 &  B \\ 
J1959+6508 & 1959+650 & J2000.0+6508 & 1ES 1959+650& 0.047& \cite{SSP93} & B& HSP& 2 &  G \\ 
J2011$-$1546 & 2008$-$159 &\n & PKS 2008$-$159& 1.180& \cite{1979ApJ...232..400P} & Q& ISP& 1 &  R \\ 
J2022+6136 & 2021+614 &\n & OW 637& 0.227& \cite{1991ApJS...75..297H} & G& LSP& 1 &  R \\ 
J2025$-$0735 & 2022$-$077 & J2025.6$-$0735 & PKS 2023$-$07& 1.388& \cite{DWF97} & Q& LSP& 2 &  G \\ 
J2031+1219 & 2029+121 & J2031.5+1219 & PKS 2029+121& 1.213& \cite{Shaw2011} & Q& LSP& 11 &  R \\ 
J2123+0535 & 2121+053 &\n & PKS 2121+053& 1.941& \cite{1991ApJ...382..433S} & Q& ISP& 1 &  R \\ 
J2131$-$1207 & 2128$-$123 &\n & PKS 2128$-$12& 0.501& \cite{1968ApJ...154L.101S} & Q& ISP& 1 &  R \\ 
J2134$-$0153 & 2131$-$021 & J2134.0$-$0203 & 4C$-$02.81& 1.284& \cite{1LAC} & Q& LSP& 1 &  R \\ 
J2136+0041 & 2134+004 &\n & PKS 2134+004& 1.932& \cite{1994ApJ...436..678O} & Q& LSP& 1 &  R \\ 
J2139+1423 & 2136+141 &\n & OX 161& 2.427& \cite{1974ApJ...190..271W} & Q& LSP& 7 &  R \\ 
J2143+1743 & 2141+175 & J2143.4+1742 & OX 169& 0.2107& \cite{HK09} & Q& ISP& 2 &  G \\ 
J2147+0929 & 2144+092 & J2147.2+0929 & PKS 2144+092& 1.113& \cite{1988ApJ...327..561W} & Q& LSP& 2 &  G \\ 
J2148+0657 & 2145+067 & J2148.5+0654 & 4C +06.69& 0.999& \cite{1991ApJ...382..433S} & Q& LSP& 1 &  R \\ 
J2158$-$1501 & 2155$-$152 & J2157.9$-$1503 & PKS 2155$-$152& 0.672& \cite{1988ApJ...327..561W} & Q& LSP& 1 &  R \\ 
J2202+4216 & 2200+420 & J2202.8+4216 & BL Lac& 0.0686& \cite{1995ApJ...452L...5V} & B& LSP& 1 &  B \\ 
J2203+1725 & 2201+171 & J2203.5+1726 & PKS 2201+171& 1.076& \cite{1977ApJ...215..427S} & Q& ISP& 1 &  G \\ 
J2203+3145 & 2201+315 &\n & 4C +31.63& 0.2947& \cite{1996ApJS..104...37M} & Q& ISP& 1 &  R \\ 
J2218$-$0335 & 2216$-$038 &\n & PKS 2216$-$03& 0.901& \cite{1967ApJ...147..837L} & Q& ISP& 1 &  R \\ 
J2225$-$0457 & 2223$-$052 & J2225.8$-$0457 & 3C 446& 1.404& \cite{WAA83} & Q& LSP& 4 &  B \\ 
J2229$-$0832 & 2227$-$088 & J2229.7$-$0832 & PHL 5225& 1.5595& \cite{2004SDSS2.C...0000:} & Q& LSP& 1 &  B \\ 
J2232+1143 & 2230+114 & J2232.5+1144 & CTA 102& 1.037& \cite{1994ApJS...93..125F} & Q& ISP& 1 &  B \\ 
J2236$-$1433 & 2233$-$148 & J2236.4$-$1432 & OY -156& \n& \n & B& LSP& 11 &  G \\ 
J2236+2828 & 2234+282 & J2236.2+2828 & CTD 135& 0.795& \cite{1991MNRAS.250..414J} & Q& LSP& 7 &  G \\ 
J2243+2021 & 2241+200 & J2244.0+2021 & RGB J2243+203& \n& \n & B& ISP& 5 &  G \\ 
J2246$-$1206 & 2243$-$123 &\n & PKS 2243$-$123& 0.632& \cite{1975MNRAS.173p..87B} & Q& ISP& 1 &  R \\ 
J2250$-$2806 & 2247$-$283 & J2250.8$-$2809 & PMN J2250$-$2806& 0.525& \cite{Shaw2011} & Q& LSP& 11 &  G \\ 
J2253+1608 & 2251+158 & J2253.9+1608 & 3C 454.3& 0.859& \cite{1991MNRAS.250..414J} & Q& ISP& 1 &  B \\ 
J2327+0940 & 2325+093 & J2327.7+0943 & OZ 042& 1.841& \cite{Shaw2011} & Q& LSP& 2 &  B \\ 
J2331$-$2148 & 2328$-$221 & J2331.0$-$2145 & PMN J2331$-$2148& 0.563& \cite{Shaw2011} & Q& \n& \n &  G \\ 
J2348$-$1631 & 2345$-$167 & J2348.0$-$1629 & PKS 2345$-$16& 0.576& \cite{TMD93} & Q& LSP& 1 &  R \\ 
\enddata 
\tablecomments{Columns are as follows: 
(1) IAU name (J2000), 
(2) IAU name (B1950), 
(3) 1FGL catalog name,
(4) other name,
(5) redshift,
(6) literature reference for redshift,
(7) optical classification, where B = BL Lac, Q=Quasar, G = Radio galaxy, and U = Unidentified,
(8) spectral energy distribution class, where HSP = high spectral peaked, ISP = intermediate spectral peaked,
and LSP = low spectral peaked.
(9) literature reference for SED data, where 
1 = \cite{Chang_thesis},
2 = \cite{abdo_sed},
3 = \cite{1FGL},
4 = Meyer et al., 2011, ApJ, submitted,
5 = \cite{Nieppola2006},
6 = \cite{Nieppola2008},
7 = \cite{Aatrokoski2011},
8 = \cite{Tavecchio2010},
9 = \cite{Ruger2010},
10 = \cite{1988AJ.....95..307I},
11 = 2LAC catalog, Abdo et al., in preparation,
(10) Sample membership, where G=1FM $\gamma$-ray selected sample, R = 1FM-matching radio sample, B = in both samples.
}

\end{deluxetable*} 

\begin{deluxetable*}{lllrrrr} 
\tablecolumns{7} 
%\tabletypesize{\scriptsize} 
\tablewidth{0pt}  
\tablecaption{\label{2fluxtable} Flux Data  }  
\tablehead{ \colhead{J2000}  &\colhead{B1950} &  
\colhead{VLBA} &\colhead{VLBA} & \colhead{Single Dish} & \colhead{Arcsecond}  & \colhead{$G_r$}  \\ 
\colhead{Name} & \colhead{Name}  & \colhead{Epoch} & \colhead{Total} & \colhead{Median} &\colhead{Emission}  &\colhead{} \\ 
& && \colhead{(Jy)} &\colhead{(Jy)} &\colhead{(Jy)}& \colhead{}  \\ 
\colhead{(1)} & \colhead{(2)} & \colhead{(3)} & \colhead{(4)} &  
\colhead{(5)} & \colhead{(6)}  & \colhead{(7)}  } 
\startdata 
J0006$-$0623 & 0003$-$066 &  2009 May 2 &  2.50&   2.41\phantom{$\;$}& \n  & $ < $   6.7  \\ 
J0017$-$0512 & 0015$-$054 &  2009 Jul 5 &  0.29&   0.32\phantom{$\;$}& \n  & $  $   972  \\ 
J0050$-$0929 & 0048$-$097 &  2008 Oct 3 &  1.09&   1.34\phantom{$\;$}& \n  & $  $   344  \\ 
J0108+0135 & 0106+013 &  2009 Jun 25 &  2.77&   2.66\phantom{$\;$}& \n  & $  $   1174  \\ 
J0112+2244 & 0109+224 &  2009 Jul 5 &  0.48&   0.79\phantom{$\;$}& \n  & $  $   489  \\ 
J0112+3208 & 0110+318 &  2009 Jun 3 &  0.70&   \n\phantom{$\;$}& \n  & $  $   1332  \\ 
J0118$-$2141 & 0116$-$219 &  2009 Jul 23 &  0.70&   \n\phantom{$\;$}& \n  & $  $   1047  \\ 
J0120$-$2701 & 0118$-$272 &  2009 Dec 26 &  0.56&   \n\phantom{$\;$}& \n  & $  $   529  \\ 
J0121+1149 & 0119+115 &  2009 Jun 15 &  3.57&   3.76\phantom{$\;$}& \n  & $ < $   9.9  \\ 
J0132$-$1654 & 0130$-$171 &  2009 Oct 27 &  2.02&   2.02\phantom{$\;$}& \n  & $  $   352  \\ 
J0136+3906 & 0133+388 &  2010 Nov 29 &  0.05&   \n\phantom{$\;$}& \n  & $  $   9763  \\ 
J0136+4751 & 0133+476 &  2009 Jun 25 &  4.44&   3.87\phantom{$\;$}& \n  & $  $   415  \\ 
J0145$-$2733 & 0142$-$278 &  2009 Dec 26 &  0.95&   \n\phantom{$\;$}& \n  & $  $   972  \\ 
J0205+3212 & 0202+319 &  2008 Aug 25 &  3.17&   3.26\phantom{$\;$}& \n  & $  $   106  \\ 
J0204$-$1701 & 0202$-$172 &  2009 Jul 5 &  1.45&   1.47\phantom{$\;$}& \n  & $  $   370  \\ 
J0217+7349 & 0212+735 &  2008 Sep 12 &  3.78&   3.72\phantom{$\;$}& \n  & $  $   296  \\ 
J0217+0144 & 0215+015 &  2008 Nov 19 &  2.00&   1.53\phantom{$\;$}& \n  & $  $   788  \\ 
J0222+4302 & 0219+428 &  2009 Jun 15 &  0.60&   0.86\phantom{$\;$}& 0.25  & $  $   3827  \\ 
J0231+1322 & 0229+131 &  2010 Oct 25 &  1.90&   1.57\phantom{$\;$}& \n  & $ < $   51  \\ 
J0237+2848 & 0234+285 &  2009 Jun 25 &  2.54&   3.14\phantom{$\;$}& \n  & $  $   427  \\ 
J0238+1636 & 0235+164 &  2009 Mar 25 &  3.08&   3.15\phantom{$\;$}& \n  & $  $   1396  \\ 
J0252$-$2219 & 0250$-$225 &  2009 Mar 25 &  0.51&   \n\phantom{$\;$}& \n  & $  $   2677  \\ 
J0303$-$2407 & 0301$-$243 &  2010 Mar 1 &  0.21&   \n\phantom{$\;$}& \n  & $  $   1933  \\ 
J0316+0904 & 0313+085 &  2010 Nov 20 &  0.06&   \n\phantom{$\;$}& \n  & $  $   4960  \\ 
J0319+4130 & 0316+413 &  2009 May 28 &  19.40&   18.91\phantom{$\;$}& \n  & $  $   63  \\ 
J0339$-$0146 & 0336$-$019 &  2009 May 2 &  2.36&   2.35\phantom{$\;$}& \n  & $  $   104  \\ 
J0349$-$2102 & 0347$-$211 &  2009 Jul 5 &  0.62&   \n\phantom{$\;$}& \n  & $  $   3981  \\ 
J0403+2600 & 0400+258 &  2010 Oct 15 &  1.85&   1.85\phantom{$\;$}& \n  & $ < $   75  \\ 
J0423$-$0120 & 0420$-$014 &  2009 Jul 5 &  6.29&   4.45\phantom{$\;$}& \n  & $  $   254  \\ 
J0433+0521 & 0430+052 &  2009 Jul 5 &  2.69&   3.18\phantom{$\;$}& 0.56  & $ < $   27  \\ 
J0433+2905 & 0430+289 &  2009 Jul 23 &  0.31&   0.30\phantom{$\;$}& \n  & $  $   1280  \\ 
J0442$-$0017 & 0440$-$003 &  2009 May 28 &  1.26&   1.24\phantom{$\;$}& \n  & $  $   1050  \\ 
J0453$-$2807 & 0451$-$282 &  2009 Aug 19 &  1.71&   \n\phantom{$\;$}& \n  & $  $   1201  \\ 
J0457$-$2324 & 0454$-$234  &  2009 Jun 25 &  1.99&   $ 1.89^a$& \n & $  $    2566   \\ 
J0507+6737 & 0502+675 &  2010 Nov 20 &  0.05&   0.03\phantom{$\;$}& \n  & $  $   9048  \\ 
J0509+0541 & 0506+056 &  2009 Jun 3 &  0.59&   0.60\phantom{$\;$}& \n  & $  $   646  \\ 
J0530+1331 & 0528+134 &  2009 Mar 25 &  2.86&   2.98\phantom{$\;$}& \n  & $  $   838  \\ 
J0532+0732 & 0529+075 &  2009 May 2 &  1.47&   1.42\phantom{$\;$}& \n  & $  $   608  \\ 
J0608$-$1520 & 0605$-$153 &  2010 Mar 1 &  0.20&   0.21\phantom{$\;$}& \n  & $  $   4471  \\ 
J0609$-$1542 & 0607$-$157 &  2009 Jun 25 &  5.17&   4.92\phantom{$\;$}& \n  & $ < $   12  \\ 
J0612+4122 & 0609+413 &  2009 Dec 26 &  0.22&   0.28\phantom{$\;$}& \n  & $  $   1022  \\ 
J0630$-$2406 & 0628$-$240 &  2010 Nov 29 &  0.07&   \n\phantom{$\;$}& \n  & $  $   4221  \\ 
J0646+4451 & 0642+449 &  2009 May 28 &  3.62&   3.43\phantom{$\;$}& \n  & $ < $   57  \\ 
J0654+4514 & 0650+453 &  2009 Jun 25 &  0.38&   0.50\phantom{$\;$}& \n  & $  $   2063  \\ 
J0654+5042 & 0650+507 &  2009 Jul 5 &  0.20&   0.23\phantom{$\;$}& \n  & $  $   2805  \\ 
J0713+1935 & 0710+196 &  2009 Aug 19 &  0.44&   \n\phantom{$\;$}& \n  & $  $   1885  \\ 
J0719+3307 & 0716+332 &  2009 Feb 25 &  0.57&   0.58\phantom{$\;$}& \n  & $  $   1193  \\ 
J0721+7120 & 0716+714 &  2009 Jun 15 &  1.20&   2.09\phantom{$\;$}& \n  & $  $   534  \\ 
J0738+1742 & 0735+178 &  2009 Jun 25 &  0.62&   0.74\phantom{$\;$}& 0.19  & $  $   629  \\ 
J0739+0137 & 0736+017 &  2009 Jul 5 &  1.20&   1.33\phantom{$\;$}& 0.20  & $  $   328  \\ 
J0748+2400 & 0745+241 &  2010 Oct 25 &  1.15&   1.54\phantom{$\;$}& \n  & $ < $   16  \\ 
J0750+1231 & 0748+126 &  2009 Feb 25 &  4.30&   4.33\phantom{$\;$}& \n  & $  $   70  \\ 
J0808$-$0751 & 0805$-$077 &  2009 Jun 25 &  1.91&   1.08\phantom{$\;$}& \n  & $  $   1835  \\ 
J0818+4222 & 0814+425 &  2009 May 28 &  1.68&   1.44\phantom{$\;$}& \n  & $  $   523  \\ 
J0825+0309 & 0823+033 &  2009 Jul 5 &  0.98&   1.53\phantom{$\;$}& \n  & $  $   70  \\ 
J0830+2410 & 0827+243 &  2008 Nov 19 &  1.53&   1.49\phantom{$\;$}& \n  & $  $   353  \\ 
J0836$-$2016 & 0834$-$201 &  2009 Mar 25 &  2.07&   \n\phantom{$\;$}& 0.65  & $ < $   118  \\ 
J0841+7053 & 0836+710 &  2009 May 2 &  1.58&   1.57\phantom{$\;$}& \n  & $  $   1028  \\ 
J0854+2006 & 0851+202 &  2009 May 28 &  4.67&   3.78\phantom{$\;$}& \n  & $  $   88  \\ 
J0909+0121 & 0906+015 &  2009 May 28 &  1.54&   1.35\phantom{$\;$}& \n  & $  $   781  \\ 
J0920+4441 & 0917+449 &  2009 Jun 25 &  2.12&   2.02\phantom{$\;$}& \n  & $  $   2154  \\ 
J0927+3902 & 0923+392 &  2009 Jul 5 &  10.86&   10.18\phantom{$\;$}& \n  & $ < $   2.4  \\ 
J0948+4039 & 0945+408 &  2009 Jun 3 &  1.69&   1.76\phantom{$\;$}& \n  & $ < $   44  \\ 
J0948+0022 & 0946+006 &  2009 May 28 &  0.44&   0.24\phantom{$\;$}& \n  & $  $   2901  \\ 
J0957+5522 & 0954+556 &  2009 Mar 25 &  0.15&   1.19\phantom{$\;$}& 0.96  & $  $   5909  \\ 
J0958+6533 & 0954+658 &  2009 Jul 5 &  1.34&   1.28\phantom{$\;$}& \n  & $  $   74  \\ 
J1012+2439 & 1009+245 &  2010 Nov 29 &  0.05&   0.05\phantom{$\;$}& \n  & $  $   14584  \\ 
J1015+4926 & 1011+496 &  2009 May 2 &  0.20&   0.28\phantom{$\;$}& 0.08  & $  $   3431  \\ 
J1016+0513 & 1013+054 &  2009 Jun 15 &  0.66&   0.62\phantom{$\;$}& \n  & $  $   2730  \\ 
J1037+5711 & 1034+574 &  2010 Mar 1 &  0.11&   0.17\phantom{$\;$}& \n  & $  $   1649  \\ 
J1037$-$2934 & 1034$-$293 &  2010 Oct 15 &  1.44&   \n\phantom{$\;$}& \n  & $ < $   13  \\ 
J1038+0512 & 1036+054 &  2008 Oct 3 &  1.49&   1.38\phantom{$\;$}& \n  & $ < $   17  \\ 
J1058+0133 & 1055+018 &  2008 Aug 25 &  4.32&   4.65\phantom{$\;$}& \n  & $  $   265  \\ 
J1058+5628 & 1055+567 &  2009 Aug 19 &  0.18&   0.17\phantom{$\;$}& \n  & $  $   3011  \\ 
J1104+3812 & 1101+384 &  2009 Jun 25 &  0.33&   0.44\phantom{$\;$}& 0.11  & $  $   6456  \\ 
J1121$-$0553 & 1118$-$056 &  2009 Jun 15 &  0.48&   \n\phantom{$\;$}& \n  & $  $   1495  \\ 
J1127$-$1857 & 1124$-$186 &  2009 May 2 &  1.74&   1.64\phantom{$\;$}& \n  & $  $   334  \\ 
J1130$-$1449 & 1127$-$145 &  2009 Jul 5 &  2.33&   2.27\phantom{$\;$}& \n  & $  $   528  \\ 
J1159+2914 & 1156+295 &  2009 Jun 3 &  2.18&   3.07\phantom{$\;$}& \n  & $  $   280  \\ 
J1215$-$1731 & 1213$-$172 &  2008 Sep 12 &  1.75&   1.80\phantom{$\;$}& 0.16  & $ < $   41  \\ 
J1217+3007 & 1215+303 &  2009 Jun 15 &  0.36&   0.38\phantom{$\;$}& \n  & $  $   1223  \\ 
J1221+3010 & 1218+304 &  2010 Nov 20 &  0.07&   \n\phantom{$\;$}& \n  & $  $   4114  \\ 
J1221+2813 & 1219+285 &  2009 May 28 &  0.33&   0.40\phantom{$\;$}& 0.07  & $  $   1837  \\ 
J1224+2122 & 1222+216 &  2009 May 28 &  1.01&   1.15\phantom{$\;$}& 0.13  & $  $   359  \\ 
J1229+0203 & 1226+023 &  2009 Jun 25 &  24.38&   27.84\phantom{$\;$}& 6.58  & $  $   83  \\ 
J1230+1223 & 1228+126 &  2009 Jul 5 &  2.51&   26.30\phantom{$\;$}& 23.71  & $  $   45  \\ 
J1239+0443 & 1236+049 &  2009 Jun 3 &  0.36&   0.38\phantom{$\;$}& \n  & $  $   2926  \\ 
J1246$-$2547 & 1244$-$255 &  2009 Jun 15 &  1.10&   \n\phantom{$\;$}& \n  & $  $   970  \\ 
J1248+5820 & 1246+586 &  2009 Oct 27 &  0.12&   0.16\phantom{$\;$}& \n  & $  $   2929  \\ 
J1256$-$0547 & 1253$-$055 &  2009 Jun 25 &  12.01&   13.65\phantom{$\;$}& \n  & $  $   328  \\ 
J1303+2433 & 1300+248 &  2010 Nov 13 &  0.11&   0.28\phantom{$\;$}& \n  & $  $   1049  \\ 
J1310+3220 & 1308+326 &  2009 Jun 3 &  2.22&   1.75\phantom{$\;$}& \n  & $  $   705  \\ 
J1332$-$0509 & 1329$-$049 &  2009 Jul 5 &  1.12&   0.99\phantom{$\;$}& \n  & $  $   3117  \\ 
J1332$-$1256 & 1329$-$126 &  2010 Mar 1 &  0.35&   \n\phantom{$\;$}& \n  & $  $   3917  \\ 
J1337$-$1257 & 1334$-$127 &  2009 Jun 25 &  6.59&   6.51\phantom{$\;$}& \n  & $  $   66  \\ 
J1344$-$1723 & 1341$-$171 &  2009 Jun 25 &  0.33&   0.39\phantom{$\;$}& \n  & $  $   3486  \\ 
J1427+2348 & 1424+240 &  2009 Jun 25 &  0.18&   0.26\phantom{$\;$}& 0.06  & $  $   5450  \\ 
J1436+6336 & 1435+638 &  2010 Jul 12 &  1.54&   1.50\phantom{$\;$}& \n  & $ < $   41  \\ 
J1504+1029 & 1502+106 &  2009 Mar 25 &  3.15&   2.65\phantom{$\;$}& \n  & $  $   5965  \\ 
J1512$-$0905 & 1510$-$089 &  2009 Jul 5 &  3.98&   2.75\phantom{$\;$}& \n  & $  $   2335  \\ 
J1516+1932 & 1514+197 &  2010 Sep 27 &  0.90&   1.56\phantom{$\;$}& \n  & $  $   64  \\ 
J1517$-$2422 & 1514$-$241 &  2009 Jun 3 &  2.32&   \n\phantom{$\;$}& \n  & $  $   168  \\ 
J1522+3144 & 1520+319 &  2009 Jun 15 &  0.42&   0.40\phantom{$\;$}& \n  & $  $   12312  \\ 
J1542+6129 & 1542+616 &  2010 Nov 29 &  0.14&   0.13\phantom{$\;$}& \n  & $  $   3628  \\ 
J1549+0237 & 1546+027 &  2009 Jun 25 &  1.79&   1.72\phantom{$\;$}& \n  & $  $   191  \\ 
J1550+0527 & 1548+056 &  2009 Jan 30 &  2.64&   2.83\phantom{$\;$}& \n  & $  $   56  \\ 
J1553+1256 & 1551+130 &  2009 Jun 3 &  0.67&   0.67\phantom{$\;$}& \n  & $  $   2032  \\ 
J1555+1111 & 1553+113 &  2009 Jun 15 &  0.15&   0.18\phantom{$\;$}& \n  & $  $   8397  \\ 
J1613+3412 & 1611+343 &  2009 May 2 &  2.81&   2.83\phantom{$\;$}& \n  & $  $   46  \\ 
J1625$-$2527 & 1622$-$253 &  2009 Oct 27 &  2.32&   \n\phantom{$\;$}& \n  & $  $   286  \\ 
J1635+3808 & 1633+382 &  2009 May 2 &  2.88&   2.80\phantom{$\;$}& \n  & $  $   933  \\ 
J1638+5720 & 1637+574 &  2009 Mar 25 &  1.81&   1.80\phantom{$\;$}& \n  & $ < $   13  \\ 
J1640+3946 & 1638+398 &  2009 May 28 &  0.78&   0.79\phantom{$\;$}& \n  & $ < $   418  \\ 
J1642+3948 & 1641+399 &  2009 Jul 5 &  9.14&   7.72\phantom{$\;$}& \n  & $  $   130  \\ 
J1642+6856 & 1642+690 &  2008 Nov 26 &  3.84&   4.62\phantom{$\;$}& \n  & $ < $   7.3  \\ 
J1653+3945 & 1652+398 &  2009 Jun 15 &  0.87&   1.17\phantom{$\;$}& 0.30  & $  $   812  \\ 
J1658+0741 & 1655+077 &  2009 Jul 5 &  1.85&   \n\phantom{$\;$}& \n  & $ < $   29  \\ 
J1700+6830 & 1700+685 &  2009 Jul 5 &  0.25&   0.30\phantom{$\;$}& \n  & $  $   1201  \\ 
J1719+1745 & 1717+178 &  2009 Jul 5 &  0.58&   0.58\phantom{$\;$}& \n  & $  $   533  \\ 
J1725+1152 & 1722+119 &  2010 Nov 20 &  0.07&   0.07\phantom{$\;$}& \n  & $  $   5390  \\ 
J1727+4530 & 1726+455 &  2008 Aug 25 &  1.02&   1.39\phantom{$\;$}& \n  & $  $   215  \\ 
J1733$-$1304 & 1730$-$130 &  2009 Jun 25 &  4.00&   4.74\phantom{$\;$}& \n  & $  $   113  \\ 
J1734+3857 & 1732+389 &  2009 Dec 26 &  0.97&   0.88\phantom{$\;$}& \n  & $  $   1097  \\ 
J1740+5211 & 1739+522 &  2008 Aug 25 &  0.94&   1.16\phantom{$\;$}& \n  & $  $   1504  \\ 
J1743$-$0350 & 1741$-$038 &  2008 Nov 19 &  3.26&   3.02\phantom{$\;$}& \n  & $ < $   33  \\ 
J1751+0939 & 1749+096 &  2009 Jun 3 &  4.20&   5.13\phantom{$\;$}& \n  & $  $   136  \\ 
J1753+2848 & 1751+288 &  2009 Jun 25 &  1.48&   1.57\phantom{$\;$}& \n  & $ < $   44  \\ 
J1801+4404 & 1800+440 &  2008 Aug 25 &  1.32&   1.44\phantom{$\;$}& \n  & $ < $   52  \\ 
J1800+7828 & 1803+784 &  2009 Mar 25 &  2.40&   2.31\phantom{$\;$}& \n  & $  $   212  \\ 
J1806+6949 & 1807+698 &  2009 Jul 5 &  1.37&   1.60\phantom{$\;$}& 0.23  & $  $   163  \\ 
J1824+5651 & 1823+568 &  2009 May 28 &  1.59&   1.61\phantom{$\;$}& \n  & $  $   266  \\ 
J1829+4844 & 1828+487  &  2009 Mar 25 &  1.80&   $ 2.81^a$& 1.27 & $  $    56   \\ 
J1842+6809 & 1842+681 &  2010 Oct 25 &  0.50&   0.88\phantom{$\;$}& \n  & $ < $   59  \\ 
J1848+3219 & 1846+322 &  2009 Jun 3 &  0.62&   0.61\phantom{$\;$}& \n  & $  $   1036  \\ 
J1849+6705 & 1849+670 &  2008 Oct 3 &  1.88&   2.60\phantom{$\;$}& \n  & $  $   700  \\ 
J1903+5540 & 1902+556 &  2010 Nov 20 &  0.18&   0.11\phantom{$\;$}& \n  & $  $   2464  \\ 
J1911$-$2006 & 1908$-$201 &  2009 Jun 25 &  1.64&   \n\phantom{$\;$}& \n  & $  $   633  \\ 
J1923$-$2104 & 1920$-$211 &  2009 Jun 15 &  2.06&   \n\phantom{$\;$}& \n  & $  $   786  \\ 
J1924$-$2914 & 1921$-$293  &  2010 Mar 1 &  15.54&   $ 14.16^a$& \n & $  $    18   \\ 
J1927+7358 & 1928+738 &  2009 May 28 &  3.71&   3.21\phantom{$\;$}& \n  & $ < $   13  \\ 
J1954$-$1123 & 1951$-$115 &  2009 Dec 26 &  0.42&   0.32\phantom{$\;$}& \n  & $  $   1705  \\ 
J1955+5131 & 1954+513 &  2010 Oct 15 &  1.26&   1.52\phantom{$\;$}& \n  & $ < $   21  \\ 
J2000$-$1748 & 1958$-$179 &  2009 Jul 5 &  2.85&   2.54\phantom{$\;$}& \n  & $  $   221  \\ 
J1959+6508 & 1959+650 &  2009 Jun 3 &  0.22&   0.21\phantom{$\;$}& 0.03  & $  $   3021  \\ 
J2011$-$1546 & 2008$-$159 &  2008 Aug 25 &  2.04&   1.99\phantom{$\;$}& \n  & $ < $   64  \\ 
J2022+6136 & 2021+614 &  2009 Jan 30 &  2.26&   2.31\phantom{$\;$}& \n  & $ < $   11  \\ 
J2025$-$0735 & 2022$-$077 &  2009 Jun 15 &  0.95&   1.11\phantom{$\;$}& \n  & $  $   2978  \\ 
J2031+1219 & 2029+121 &  2010 Oct 15 &  1.26&   1.36\phantom{$\;$}& \n  & $  $   262  \\ 
J2123+0535 & 2121+053 &  2009 May 28 &  1.92&   1.65\phantom{$\;$}& \n  & $ < $   81  \\ 
J2131$-$1207 & 2128$-$123 &  2009 Jan 7 &  2.23&   2.27\phantom{$\;$}& \n  & $ < $   18  \\ 
J2134$-$0153 & 2131$-$021 &  2009 Feb 25 &  2.41&   2.31\phantom{$\;$}& \n  & $  $   54  \\ 
J2136+0041 & 2134+004 &  2008 Nov 19 &  6.67&   6.53\phantom{$\;$}& \n  & $ < $   14  \\ 
J2139+1423 & 2136+141 &  2009 Jul 5 &  2.71&   2.53\phantom{$\;$}& \n  & $ < $   32  \\ 
J2143+1743 & 2141+175 &  2009 Jun 3 &  1.09&   0.81\phantom{$\;$}& \n  & $  $   816  \\ 
J2147+0929 & 2144+092 &  2009 Jun 25 &  1.30&   0.83\phantom{$\;$}& \n  & $  $   1878  \\ 
J2148+0657 & 2145+067 &  2009 Mar 25 &  5.57&   5.54\phantom{$\;$}& \n  & $  $   38  \\ 
J2158$-$1501 & 2155$-$152 &  2009 May 2 &  1.69&   1.61\phantom{$\;$}& \n  & $  $   90  \\ 
J2202+4216 & 2200+420 &  2009 Jun 15 &  4.52&   3.23\phantom{$\;$}& \n  & $  $   180  \\ 
J2203+1725 & 2201+171 &  2009 Jul 5 &  1.16&   1.07\phantom{$\;$}& \n  & $  $   889  \\ 
J2203+3145 & 2201+315 &  2009 Feb 25 &  2.60&   2.57\phantom{$\;$}& \n  & $ < $   5.3  \\ 
J2218$-$0335 & 2216$-$038 &  2009 Mar 25 &  1.50&   1.60\phantom{$\;$}& \n  & $ < $   14  \\ 
J2225$-$0457 & 2223$-$052 &  2009 May 2 &  7.51&   8.05\phantom{$\;$}& \n  & $  $   96  \\ 
J2229$-$0832 & 2227$-$088 &  2009 Jun 3 &  2.75&   2.62\phantom{$\;$}& \n  & $  $   973  \\ 
J2232+1143 & 2230+114 &  2009 Mar 25 &  3.87&   5.23\phantom{$\;$}& \n  & $  $   238  \\ 
J2236$-$1433 & 2233$-$148 &  2009 Dec 26 &  0.52&   0.45\phantom{$\;$}& \n  & $  $   672  \\ 
J2236+2828 & 2234+282 &  2009 Dec 26 &  1.21&   1.22\phantom{$\;$}& \n  & $  $   607  \\ 
J2243+2021 & 2241+200 &  2010 Nov 29 &  0.07&   0.07\phantom{$\;$}& \n  & $  $   5133  \\ 
J2246$-$1206 & 2243$-$123 &  2009 Jun 15 &  2.19&   2.18\phantom{$\;$}& \n  & $ < $   23  \\ 
J2250$-$2806 & 2247$-$283 &  2009 Jun 3 &  0.51&   \n\phantom{$\;$}& \n  & $  $   781  \\ 
J2253+1608 & 2251+158 &  2009 Jun 25 &  6.83&   12.74\phantom{$\;$}& \n  & $  $   788  \\ 
J2327+0940 & 2325+093 &  2009 Jun 15 &  2.01&   2.44\phantom{$\;$}& \n  & $  $   1092  \\ 
J2331$-$2148 & 2328$-$221 &  2010 Nov 29 &  0.14&   \n\phantom{$\;$}& \n  & $  $   3278  \\ 
J2348$-$1631 & 2345$-$167 &  2009 May 2 &  2.23&   2.04\phantom{$\;$}& \n  & $  $   120  \\ 
\enddata 
\tablecomments{Columns are as follows: 
(1) IAU name (J2000),
(2) IAU name (B1950), 
(3) VLBA observation date,
(4) total 15 GHz VLBA flux density in Jy,
(5) single dish OVRO 15 GHz median flux density in Jy during 11 month Fermi era. The $a$ flag indicates UMRAO 14.5 GHz data,
(6) arcsecond scale 15 GHz flux density in Jy,
(7) ratio of average $>$ 100 MeV $\gamma$-ray energy luminosity to 15 GHz radio luminosity.
}

\end{deluxetable*} 

\begin{deluxetable}{llrrrrr} 
\tablecolumns{7} 
%\tabletypesize{\scriptsize} 
\tablewidth{0pt}  
\tablecaption{\label{3jettable} Jet Data  }  
\tablehead{ \colhead{J2000}  &\colhead{B1950} & \colhead{Op.} &\colhead{Jet} &\colhead{Core}& \colhead{$m$} & \colhead{Core}  \\ 
\colhead{Name} & \colhead{Name}  & \colhead{Angle} & \colhead{P.A.} & \colhead{$T_b$} &  \colhead{(\%)} &\colhead{EVPA} \\ 
& &\colhead{(deg)}& \colhead{(deg)} &\colhead{(K)} &\colhead{}& \colhead{(deg)}   \\ 
\colhead{(1)} & \colhead{(2)} & \colhead{(3)} & \colhead{(4)} &  
\colhead{(5)} & \colhead{(6)}  & \colhead{(7)}} 
\startdata 
J0006$-$0623 & 0003$-$066 &  22 &  $-$95 & $ > $  12.8 & $ $ 7.8 & 15  \\ 
J0017$-$0512 & 0015$-$054 &  39 &  $-$123 & $  $  11.4 & $ <$ 0.3 & \n  \\ 
J0050$-$0929 & 0048$-$097 &  15 &  $-$8 & $ > $  13.2 & $ $ 3.7 & 150  \\ 
J0108+0135 & 0106+013 &  28 &  $-$127 & $  $  12.6 & $ $ 0.9 & 113  \\ 
J0112+2244 & 0109+224 &  22 &  86 & $  $  11.3 & $ $ 1.5 & 68  \\ 
J0112+3208 & 0110+318 &  18 &  $-$66 & $  $  12.2 & $ $ 2.2 & 115  \\ 
J0118$-$2141 & 0116$-$219 &  32 &  $-$69 & $  $  11.2 & $ $ 1.1 & 122  \\ 
J0120$-$2701 & 0118$-$272 &  13 &  $-$26 & $  $  11.1 & $ $ 5.5 & 136  \\ 
J0121+1149 & 0119+115 &  15 &  3 & $  $  12.9 & $ $ 5.1 & 156  \\ 
J0132$-$1654 & 0130$-$171 &  21 &  $-$109 & $  $  12.1 & $ $ 2.5 & 0  \\ 
J0136+3906 & 0133+388 &  \n &  \n & $ > $  10.6 & $ $ \n & \n  \\ 
J0136+4751 & 0133+476 &  21 &  $-$38 & $  $  12.7 & $ $ 2.0 & 95  \\ 
J0145$-$2733 & 0142$-$278 &  25 &  54 & $  $  11.4 & $ $ 1.3 & 96  \\ 
J0205+3212 & 0202+319 &  12 &  $-$11 & $  $  12.3 & $ $ 3.5 & 108  \\ 
J0204$-$1701 & 0202$-$172 &  15 &  7 & $  $  12.5 & $ $ 3.7 & 93  \\ 
J0217+7349 & 0212+735 &  12 &  113 & $ > $  13.9 & $ $ 1.3 & 42  \\ 
J0217+0144 & 0215+015 &  47 &  108 & $  $  12.6 & $ $ 3.2 & 4  \\ 
J0222+4302 & 0219+428 &  20 &  171 & $  $  12.0 & $ $ 2.9 & 25  \\ 
J0231+1322 & 0229+131 &  27 &  64 & $ > $  13.5 & $ $ 4.1 & 179  \\ 
J0237+2848 & 0234+285 &  23 &  $-$13 & $  $  12.1 & $ $ 3.9 & 135  \\ 
J0238+1636 & 0235+164 &  19 &  $-$34 & $  $  12.0 & $ $ 0.5 & 8  \\ 
J0252$-$2219 & 0250$-$225 &  68 &  $-$155 & $ > $  12.8 & $ $ 2.4 & 16  \\ 
J0303$-$2407 & 0301$-$243 &  25 &  $-$125 & $  $  10.7 & $ $ 1.0 & 50  \\ 
J0316+0904 & 0313+085 &  21 &  24 & $  $  10.5 & $ $ \n & \n  \\ 
J0319+4130 & 0316+413 &  30 &  $-$176 & $  $  11.1 & $ $ 0.04 & 123  \\ 
J0339$-$0146 & 0336$-$019 &  33 &  61 & $  $  11.8 & $ $ 3.6 & 100  \\ 
J0349$-$2102 & 0347$-$211 &  15 &  $-$147 & $  $  12.7 & $ $ 1.8 & 31  \\ 
J0403+2600 & 0400+258 &  13 &  77 & $  $  11.2 & $ $ 4.0 & 130  \\ 
J0423$-$0120 & 0420$-$014 &  24 &  $-$161 & $  $  12.2 & $ $ 2.2 & 131  \\ 
J0433+0521 & 0430+052 &  13 &  $-$115 & $ > $  12.0 & $ <$ 0.2 & \n  \\ 
J0433+2905 & 0430+289 &  54 &  56 & $  $  11.3 & $ $ 2.6 & 41  \\ 
J0442$-$0017 & 0440$-$003 &  42 &  $-$130 & $  $  11.3 & $ $ 2.3 & 172  \\ 
J0453$-$2807 & 0451$-$282 &  9 &  8 & $  $  12.7 & $ $ 1.3 & 48  \\ 
J0457$-$2324 & 0454$-$234 &  31 &  134 & $ > $  13.3 & $ $ 1.0 & 160  \\ 
J0507+6737 & 0502+675 &  \n &  \n & $  $  10.7 & $ $ \n & \n  \\ 
J0509+0541 & 0506+056 &  26 &  $-$173 & $  $  11.1 & $ $ 1.1 & 139  \\ 
J0530+1331 & 0528+134 &  20 &  52 & $  $  12.1 & $ $ 2.7 & 166  \\ 
J0532+0732 & 0529+075 &  50 &  $-$25 & $  $  10.4 & $ $ 3.8 & 165  \\ 
J0608$-$1520 & 0605$-$153 &  56 &  100 & $  $  11.0 & $ <$ 0.5 & \n  \\ 
J0609$-$1542 & 0607$-$157 &  35 &  68 & $  $  11.2 & $ $ 4.8 & 82  \\ 
J0612+4122 & 0609+413 &  20 &  119 & $  $  11.5 & $ $ 0.5 & 178  \\ 
J0630$-$2406 & 0628$-$240 &  30 &  $-$151 & $  $  10.5 & $ $ \n & \n  \\ 
J0646+4451 & 0642+449 &  21 &  83 & $  $  12.5 & $ $ 1.6 & 164  \\ 
J0654+4514 & 0650+453 &  46 &  97 & $  $  11.9 & $ $ 0.5 & 42  \\ 
J0654+5042 & 0650+507 &  20 &  93 & $  $  10.8 & $ $ 5.0 & 101  \\ 
J0713+1935 & 0710+196 &  42 &  87 & $  $  11.9 & $ $ 1.4 & 108  \\ 
J0719+3307 & 0716+332 &  22 &  76 & $  $  12.1 & $ $ 1.8 & 99  \\ 
J0721+7120 & 0716+714 &  18 &  18 & $  $  12.7 & $ $ 2.3 & 154  \\ 
J0738+1742 & 0735+178 &  23 &  63 & $  $  11.3 & $ $ 1.6 & 129  \\ 
J0739+0137 & 0736+017 &  21 &  $-$79 & $  $  11.7 & $ $ 1.1 & 168  \\ 
J0748+2400 & 0745+241 &  15 &  $-$59 & $  $  11.9 & $ $ 2.0 & 85  \\ 
J0750+1231 & 0748+126 &  23 &  89 & $  $  12.1 & $ $ 2.6 & 35  \\ 
J0808$-$0751 & 0805$-$077 &  20 &  $-$30 & $  $  13.1 & $ $ 1.9 & 154  \\ 
J0818+4222 & 0814+425 &  41 &  100 & $  $  12.2 & $ $ 1.6 & 2  \\ 
J0825+0309 & 0823+033 &  24 &  26 & $  $  12.6 & $ $ 5.3 & 41  \\ 
J0830+2410 & 0827+243 &  21 &  124 & $  $  11.9 & $ $ 2.3 & 25  \\ 
J0836$-$2016 & 0834$-$201 &  34 &  $-$100 & $  $  10.4 & $ $ 1.6 & 136  \\ 
J0841+7053 & 0836+710 &  10 &  $-$145 & $  $  12.6 & $ $ 0.1 & 93  \\ 
J0854+2006 & 0851+202 &  29 &  $-$115 & $  $  12.4 & $ $ 5.9 & 156  \\ 
J0909+0121 & 0906+015 &  19 &  43 & $  $  12.2 & $ $ 2.5 & 130  \\ 
J0920+4441 & 0917+449 &  17 &  178 & $  $  12.7 & $ $ 3.7 & 119  \\ 
J0927+3902 & 0923+392 &  16 &  101 & $  $  10.6 & $ <$ 0.7 & \n  \\ 
J0948+4039 & 0945+408 &  17 &  116 & $  $  12.1 & $ $ 2.1 & 3  \\ 
J0948+0022 & 0946+006 &  21 &  24 & $ > $  12.8 & $ $ 0.8 & 142  \\ 
J0957+5522 & 0954+556 &  \n &  \n & $  $  8.5 & $ $ 6.9 & 9  \\ 
J0958+6533 & 0954+658 &  30 &  $-$38 & $  $  11.9 & $ $ 2.4 & 51  \\ 
J1012+2439 & 1009+245 &  23 &  38 & $  $  10.7 & $ $ \n & \n  \\ 
J1015+4926 & 1011+496 &  20 &  $-$105 & $  $  11.3 & $ $ 1.1 & 134  \\ 
J1016+0513 & 1013+054 &  28 &  140 & $  $  12.2 & $ $ 2.5 & 97  \\ 
J1037+5711 & 1034+574 &  \n &  $-$167 & $  $  10.7 & $ <$ 0.8 & \n  \\ 
J1037$-$2934 & 1034$-$293 &  32 &  123 & $  $  11.6 & $ $ 3.4 & 23  \\ 
J1038+0512 & 1036+054 &  12 &  $-$5 & $  $  12.6 & $ $ 6.8 & 154  \\ 
J1058+0133 & 1055+018 &  28 &  $-$55 & $  $  12.2 & $ $ 6.4 & 127  \\ 
J1058+5628 & 1055+567 &  36 &  $-$85 & $  $  10.7 & $ <$ 0.5 & \n  \\ 
J1104+3812 & 1101+384 &  27 &  $-$34 & $ > $  12.4 & $ $ 1.3 & 94  \\ 
J1121$-$0553 & 1118$-$056 &  11 &  31 & $  $  12.0 & $ $ 2.0 & 139  \\ 
J1127$-$1857 & 1124$-$186 &  15 &  169 & $  $  12.5 & $ $ 2.0 & 106  \\ 
J1130$-$1449 & 1127$-$145 &  18 &  81 & $  $  11.9 & $ $ 0.5 & 38  \\ 
J1159+2914 & 1156+295 &  20 &  9 & $  $  12.1 & $ $ 1.8 & 16  \\ 
J1215$-$1731 & 1213$-$172 &  23 &  112 & $  $  11.3 & $ $ 3.2 & 83  \\ 
J1217+3007 & 1215+303 &  13 &  144 & $  $  11.4 & $ <$ 0.2 & \n  \\ 
J1221+3010 & 1218+304 &  22 &  94 & $  $  10.5 & $ $ \n & \n  \\ 
J1221+2813 & 1219+285 &  16 &  112 & $  $  11.6 & $ $ 1.3 & 2  \\ 
J1224+2122 & 1222+216 &  13 &  $-$2 & $  $  11.8 & $ $ 6.4 & 8  \\ 
J1229+0203 & 1226+023 &  12 &  $-$125 & $  $  12.1 & $ $ 0.2 & 10  \\ 
J1230+1223 & 1228+126 &  13 &  $-$73 & $  $  10.9 & $ $ 0.1 & 0  \\ 
J1239+0443 & 1236+049 &  29 &  $-$60 & $  $  12.3 & $ $ 1.1 & 100  \\ 
J1246$-$2547 & 1244$-$255 &  22 &  140 & $ > $  13.1 & $ $ 1.3 & 50  \\ 
J1248+5820 & 1246+586 &  47 &  4 & $  $  11.1 & $ <$ 0.7 & \n  \\ 
J1256$-$0547 & 1253$-$055 &  16 &  $-$124 & $  $  12.9 & $ $ 2.0 & 65  \\ 
J1303+2433 & 1300+248 &  \n &  $-$41 & $  $  11.7 & $ <$ 0.5 & \n  \\ 
J1310+3220 & 1308+326 &  38 &  $-$59 & $  $  12.2 & $ $ 1.4 & 77  \\ 
J1332$-$0509 & 1329$-$049 &  14 &  18 & $  $  12.7 & $ <$ 0.07 & \n  \\ 
J1332$-$1256 & 1329$-$126 &  25 &  112 & $ > $  12.5 & $ $ 0.7 & 87  \\ 
J1337$-$1257 & 1334$-$127 &  19 &  149 & $  $  12.6 & $ $ 4.3 & 169  \\ 
J1344$-$1723 & 1341$-$171 &  53 &  $-$56 & $ > $  12.7 & $ $ 1.6 & 21  \\ 
J1427+2348 & 1424+240 &  56 &  145 & $  $  11.0 & $ $ 2.1 & 153  \\ 
J1436+6336 & 1435+638 &  5 &  $-$127 & $  $  10.7 & $ <$ 0.5 & \n  \\ 
J1504+1029 & 1502+106 &  43 &  116 & $  $  13.1 & $ $ 1.3 & 164  \\ 
J1512$-$0905 & 1510$-$089 &  19 &  $-$32 & $  $  12.7 & $ $ 2.3 & 151  \\ 
J1516+1932 & 1514+197 &  19 &  $-$24 & $  $  12.6 & $ $ 2.3 & 168  \\ 
J1517$-$2422 & 1514$-$241 &  10 &  161 & $  $  11.1 & $ $ 0.6 & 91  \\ 
J1522+3144 & 1520+319 &  63 &  14 & $  $  11.4 & $ $ 1.4 & 59  \\ 
J1542+6129 & 1542+616 &  14 &  109 & $  $  11.4 & $ $ 1.8 & 139  \\ 
J1549+0237 & 1546+027 &  16 &  175 & $ > $  13.3 & $ $ 2.8 & 46  \\ 
J1550+0527 & 1548+056 &  14 &  $-$6 & $  $  12.1 & $ $ 4.4 & 141  \\ 
J1553+1256 & 1551+130 &  14 &  11 & $  $  12.2 & $ $ 1.8 & 70  \\ 
J1555+1111 & 1553+113 &  45 &  48 & $  $  10.7 & $ <$ 0.5 & \n  \\ 
J1613+3412 & 1611+343 &  28 &  168 & $  $  11.6 & $ $ 2.2 & 85  \\ 
J1625$-$2527 & 1622$-$253 &  23 &  14 & $  $  11.7 & $ $ 1.2 & 111  \\ 
J1635+3808 & 1633+382 &  21 &  $-$79 & $  $  12.8 & $ $ 0.5 & 101  \\ 
J1638+5720 & 1637+574 &  14 &  $-$156 & $  $  13.4 & $ $ 0.3 & 146  \\ 
J1640+3946 & 1638+398 &  66 &  $-$77 & $  $  11.5 & $ $ 0.8 & 141  \\ 
J1642+3948 & 1641+399 &  16 &  $-$89 & $  $  12.6 & $ $ 0.7 & 122  \\ 
J1642+6856 & 1642+690 &  15 &  $-$167 & $  $  12.7 & $ $ 4.9 & 104  \\ 
J1653+3945 & 1652+398 &  28 &  128 & $  $  11.0 & $ $ 0.5 & 105  \\ 
J1658+0741 & 1655+077 &  15 &  $-$42 & $  $  12.9 & $ $ 5.6 & 103  \\ 
J1700+6830 & 1700+685 &  17 &  142 & $ > $  12.2 & $ $ 0.5 & 38  \\ 
J1719+1745 & 1717+178 &  10 &  $-$157 & $  $  11.8 & $ $ 9.4 & 31  \\ 
J1725+1152 & 1722+119 &  \n &  \n & $ > $  10.7 & $ $ \n & \n  \\ 
J1727+4530 & 1726+455 &  26 &  $-$110 & $  $  12.6 & $ $ 1.2 & 100  \\ 
J1733$-$1304 & 1730$-$130 &  12 &  8 & $  $  12.6 & $ $ 3.0 & 58  \\ 
J1734+3857 & 1732+389 &  25 &  117 & $  $  12.3 & $ $ 2.1 & 169  \\ 
J1740+5211 & 1739+522 &  62 &  15 & $  $  12.3 & $ $ 0.7 & 77  \\ 
J1743$-$0350 & 1741$-$038 &  22 &  $-$161 & $  $  11.7 & $ $ 2.6 & 149  \\ 
J1751+0939 & 1749+096 &  28 &  17 & $  $  12.7 & $ $ 4.1 & 7  \\ 
J1753+2848 & 1751+288 &  22 &  9 & $ > $  13.1 & $ $ 0.9 & 18  \\ 
J1801+4404 & 1800+440 &  22 &  $-$156 & $  $  11.7 & $ $ 1.5 & 46  \\ 
J1800+7828 & 1803+784 &  22 &  $-$90 & $  $  12.1 & $ $ 2.5 & 77  \\ 
J1806+6949 & 1807+698 &  10 &  $-$101 & $  $  11.3 & $ <$ 0.09 & \n  \\ 
J1824+5651 & 1823+568 &  8 &  $-$160 & $  $  12.5 & $ $ 6.8 & 14  \\ 
J1829+4844 & 1828+487 &  15 &  $-$40 & $  $  12.1 & $ $ 1.3 & 90  \\ 
J1842+6809 & 1842+681 &  12 &  138 & $  $  11.9 & $ $ 2.2 & 129  \\ 
J1848+3219 & 1846+322 &  24 &  $-$41 & $ > $  13.2 & $ $ 1.7 & 144  \\ 
J1849+6705 & 1849+670 &  18 &  $-$45 & $  $  12.7 & $ $ 2.1 & 84  \\ 
J1903+5540 & 1902+556 &  32 &  41 & $  $  11.8 & $ $ 5.4 & 22  \\ 
J1911$-$2006 & 1908$-$201 &  25 &  19 & $  $  13.0 & $ $ 0.7 & 67  \\ 
J1923$-$2104 & 1920$-$211 &  30 &  $-$8 & $  $  13.4 & $ $ 1.0 & 134  \\ 
J1924$-$2914 & 1921$-$293 &  36 &  17 & $  $  12.2 & $ $ 3.0 & 131  \\ 
J1927+7358 & 1928+738 &  9 &  162 & $  $  12.0 & $ $ 0.10 & 178  \\ 
J1954$-$1123 & 1951$-$115 &  27 &  10 & $  $  11.8 & $ $ 6.4 & 123  \\ 
J1955+5131 & 1954+513 &  19 &  $-$59 & $  $  11.9 & $ $ 3.0 & 43  \\ 
J2000$-$1748 & 1958$-$179 &  24 &  105 & $  $  12.5 & $ $ 1.3 & 11  \\ 
J1959+6508 & 1959+650 &  37 &  139 & $  $  11.0 & $ $ 2.3 & 149  \\ 
J2011$-$1546 & 2008$-$159 &  14 &  12 & $  $  11.8 & $ $ 1.3 & 14  \\ 
J2022+6136 & 2021+614 &  6 &  32 & $  $  10.6 & $ $ 0.1 & 137  \\ 
J2025$-$0735 & 2022$-$077 &  19 &  $-$13 & $  $  12.2 & $ $ 2.3 & 128  \\ 
J2031+1219 & 2029+121 &  19 &  $-$154 & $  $  12.3 & $ $ 0.8 & 101  \\ 
J2123+0535 & 2121+053 &  18 &  $-$97 & $  $  11.9 & $ $ 8.1 & 25  \\ 
J2131$-$1207 & 2128$-$123 &  11 &  $-$150 & $  $  11.3 & $ $ 0.6 & 54  \\ 
J2134$-$0153 & 2131$-$021 &  35 &  104 & $  $  12.1 & $ $ 7.0 & 89  \\ 
J2136+0041 & 2134+004 &  22 &  $-$84 & $  $  12.4 & $ $ 1.9 & 22  \\ 
J2139+1423 & 2136+141 &  31 &  $-$76 & $  $  12.0 & $ $ 4.1 & 139  \\ 
J2143+1743 & 2141+175 &  31 &  $-$52 & $  $  11.7 & $ $ 1.0 & 90  \\ 
J2147+0929 & 2144+092 &  37 &  78 & $  $  12.7 & $ $ 2.0 & 11  \\ 
J2148+0657 & 2145+067 &  27 &  118 & $  $  11.8 & $ $ 0.5 & 40  \\ 
J2158$-$1501 & 2155$-$152 &  18 &  $-$148 & $  $  11.9 & $ $ 3.0 & 18  \\ 
J2202+4216 & 2200+420 &  27 &  $-$171 & $  $  12.1 & $ $ 8.1 & 13  \\ 
J2203+1725 & 2201+171 &  21 &  49 & $  $  12.5 & $ $ 0.9 & 135  \\ 
J2203+3145 & 2201+315 &  15 &  $-$144 & $  $  12.0 & $ $ 0.9 & 122  \\ 
J2218$-$0335 & 2216$-$038 &  14 &  $-$172 & $  $  11.2 & $ $ 1.2 & 157  \\ 
J2225$-$0457 & 2223$-$052 &  24 &  98 & $  $  12.5 & $ $ 2.5 & 33  \\ 
J2229$-$0832 & 2227$-$088 &  15 &  $-$10 & $  $  12.8 & $ $ 1.5 & 172  \\ 
J2232+1143 & 2230+114 &  15 &  152 & $  $  12.8 & $ $ 1.7 & 80  \\ 
J2236$-$1433 & 2233$-$148 &  42 &  105 & $  $  11.3 & $ $ 7.4 & 84  \\ 
J2236+2828 & 2234+282 &  25 &  $-$135 & $  $  10.8 & $ $ 4.8 & 36  \\ 
J2243+2021 & 2241+200 &  7 &  9 & $  $  10.7 & $ $ \n & \n  \\ 
J2246$-$1206 & 2243$-$123 &  14 &  8 & $  $  12.0 & $ $ 2.3 & 124  \\ 
J2250$-$2806 & 2247$-$283 &  20 &  159 & $ > $  12.6 & $ $ 2.0 & 23  \\ 
J2253+1608 & 2251+158 &  48 &  $-$76 & $  $  12.3 & $ $ 2.2 & 151  \\ 
J2327+0940 & 2325+093 &  32 &  $-$96 & $  $  12.6 & $ $ 1.6 & 91  \\ 
J2331$-$2148 & 2328$-$221 &  13 &  153 & $  $  11.4 & $ <$ 0.4 & \n  \\ 
J2348$-$1631 & 2345$-$167 &  28 &  124 & $  $  12.5 & $ $ 2.5 & 41  \\ 
\enddata 
\tablecomments{Columns are as follows: 
(1) IAU name (J2000),
(2) IAU name (B1950), 
(3) opening angle of the jet (degrees),
(4) position angle of the parsec-scale jet (degrees),
(5) log brightness temperature of the core (K),
(6) fractional linear polarization of the core in per cent,
(7) linear polarization electric vector position angle at the location of the core (degrees). 
}

\end{deluxetable}

\end{document}